\documentclass[longbibliography,twocolumn,showpacs,amsmath,amssymb,floatfix]{revtex4}
\usepackage{amsfonts, relsize}
\usepackage{graphics}
\usepackage{graphicx}
\usepackage{hyperref}

\usepackage{color}
\begin{document}

\title{Zero-modes and global antiferromagnetism in strained graphene}

\author{Bitan Roy}
\affiliation{ National High Magnetic Field Laboratory and Department of Physics, Florida State University, FL 32306, USA}
\affiliation{ Condensed Matter Theory Center, Department of Physics, University of Maryland, College Park, MD 20742, USA\footnote{Current affiliation}}

\author{Fakher F. Assaad}
\affiliation{Institut f\"ur Theoretische Physik und Astrophysik, Universit\"at W\"urzburg, Am Hubland, D-97074 W\"urzburg, Germany }
\affiliation{ Max-Planck-Institut f\"ur Physik komplexer Systeme, N\"othnitzer Str. 38, 01187 Dresden, Germany}

\author{Igor F. Herbut}

\affiliation{ Max-Planck-Institut f\"ur Physik komplexer Systeme, N\"othnitzer Str. 38, 01187 Dresden, Germany}
\affiliation{ Department of Physics, Simon Fraser University, Burnaby, British Columbia, Canada V5A 1S6}

\date{\today}

\begin{abstract}
A novel magnetic ground state is reported for the Hubbard Hamiltonian in strained graphene. When the chemical potential lies close to the Dirac point, the ground state exhibits locally both the N\'{e}el and ferromagnetic orders, even for weak Hubbard interaction. Whereas the N\'{e}el order parameter remains of the same sign in the entire system, the magnetization at the boundary takes the opposite sign from the bulk. The total magnetization this way vanishes, and the magnetic ground state is globally only an antiferromagnet. This peculiar ordering stems from the nature of the strain-induced single particle zero-energy states, which have support on one sublattice of the honeycomb lattice in the bulk, and on the other sublattice near the boundary of a finite system. We support our claim with the self-consistent numerical calculation of the order parameters, as well as by the Monte Carlo simulations of the Hubbard model in both uniformly and non-uniformly strained honeycomb lattice. The present result is contrasted with the magnetic ground state of the same Hubbard model in the presence of a true magnetic field (and for vanishing Zeeman coupling), which is exclusively N\'{e}el ordered, with zero local magnetization everywhere in the system.
\end{abstract}


\maketitle

\vspace{10pt}

\section{Introduction}

Despite the manifest stability of the  Dirac fermions in graphene against the effects of Coulomb interaction,\cite{Gonzales99} the nature of the possible broken symmetry phases at strong coupling continues to be an issue of fundamental importance \cite{Sorella92,Paiva05, Khveshchenko01,Khveshchenko04, Herbut06,Herbut09, Drut09,Drut09a,Drut09b, Castro11,Grushin13, Gonzales13}. The minimal onsite Hubbard interaction, for example, when sufficiently strong, is believed to produce the antiferromagnetic N\'{e}el ground state \cite{Sorella92,Paiva05, Herbut06,Herbut09, Meng10, Sorella12, Assaad13, Lang13}. The universality class of the semimetal-to-N\'{e}el insulator quantum phase transition can be captured by the effective Gross-Neveu-Yukawa field theory\cite{Herbut09a,Roy11,Rosenstein93}, and studied systematically near the upper critical (spatial) dimension of three. Recent quantum Monte Carlo simulations of the Hubbard model in a half-filled honeycomb lattice suggest a \emph{direct} transition from the Dirac semimetal to the N\'{e}el state, with the critical exponents in reasonable agreement with the field-theoretic predictions.\cite{Sorella12, Assaad13}  The correlated phases of graphene, N\'{e}el state included, unfortunately may be laying at too strong a coupling to be realized in graphene in its pristine state, even when placed in vacuum. \cite{Kotov12}

\begin{figure}[htb]
\includegraphics[width=8.75cm,height=6.0cm]{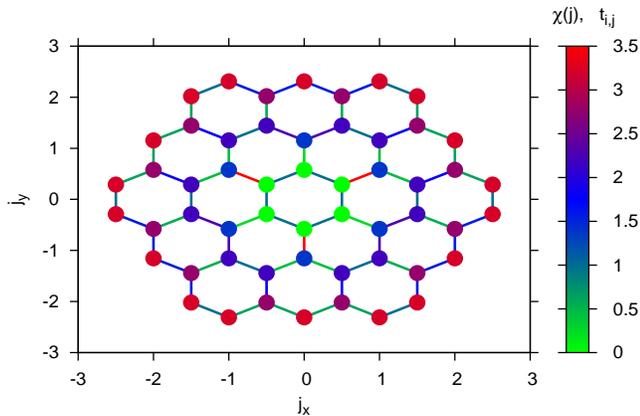}
\caption{(Color online) A realization of  strained honeycomb lattice that yields a finite axial magnetic field. Colors at each site correspond to the parameter $\chi$ in Eq.~(\ref{hoppinglattice}), which defines the modification of the hopping amplitude along each bond. }
\label{chilattice}
\end{figure}

The deformability of the graphene membrane may facilitate a different path towards the realization of some of the symmetry-broken phases at weaker couplings. \cite{Herbut08} Arguments along this line and in favor of the topological quantum anomalous (spin) Hall insulator  at weak finite-range repulsion \cite{Herbut08, Ghaemi12, Roy13}, or of unconventional superconductors\cite{Roy13a} at weak attraction, have recently been put forward. The physical reason behind this electro-mechanical phenomenon is the generic appearance of the single-particle zero-energy states in graphene under strain. Upon bulging a graphene flake, the quasi-relativistic low-energy electronic degrees of freedom couple to a  finite, time-reversal-symmetric, magnetic-like field.\cite{Guinea10}  Such an, and not necessarily uniform,  {\it axial} magnetic field, similarly to the true magnetic field, falls under the jurisdiction of index theorems,\cite{Aharonov79} and as such brings a finite number of states close to the Dirac point. This creates an ideal situation for the fermions to form various particle-hole or particle-particle condensates at weak interactions. The single-particle states at zero energy, responsible for the weak-coupling instabilities, are in the axial case, however, special:\cite{Herbut08} normalizability forces them to live exclusively on one of the two sublattices of the honeycomb lattice. The remaining states in the zero-energy subspace, which would be discarded from the spectrum as non-normalizable in an infinite system, in a finite system are found near the boundary, and to be living on the opposite sublattice.

\begin{figure}[htb]
\includegraphics[width=8.5cm,height=6.25cm]{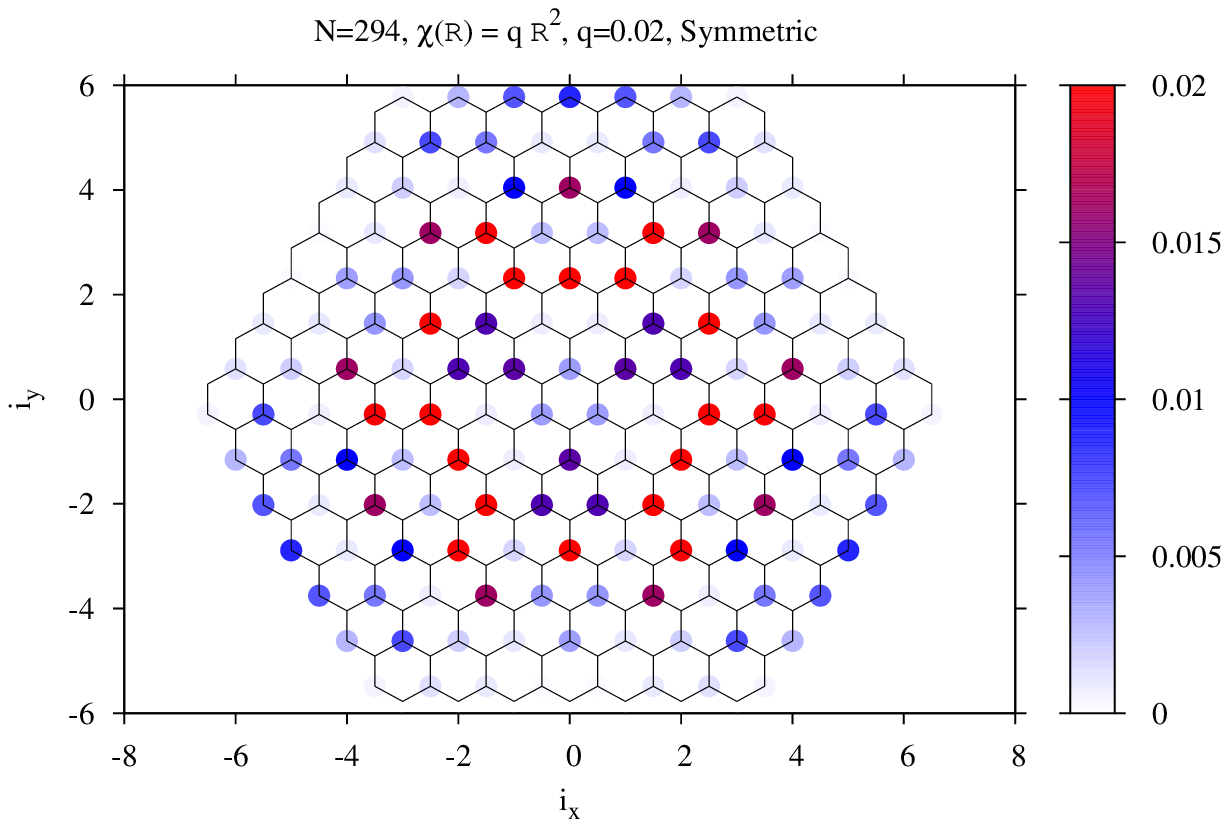}
\includegraphics[width=8.5cm,height=6.25cm]{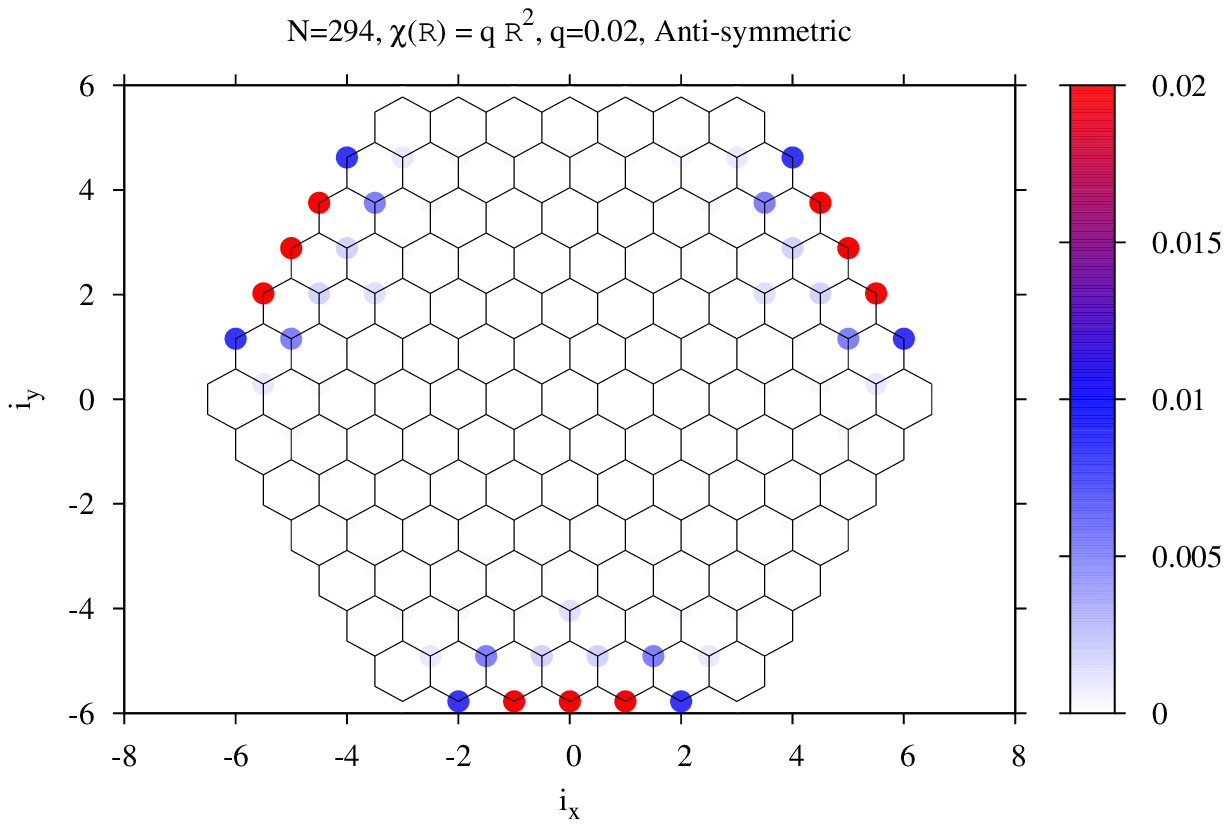}
\caption{{\color{black} (Color online) Top: spatial distribution of the symmetric combination $(|S\rangle)$ of the wave-functions $|\pm\delta \rangle$, living on A sublattice, in a system of $294$ sites or $r_{max}=7$. Spatial distribution of their anti-symmetric combination $(|AS \rangle)$, living on B sublattice, in the same system. The axial magnetic field is set to be uniform here, i.e., $\chi({\cal R})= q {\cal R}^2$, with $q=0.02$.}}
\label{zeromodes}
\end{figure}

Due to this inextricable correlation between the real-space and sublattice degrees of freedom within the zero-energy subspace the ground state of the Hubbard model in strained graphene is non-trivial. While it seems natural to expect that a short-range interaction such as Hubbard's would lead to a spin-polarized ``Hund"  ferromagnetic state in the presence of a flat band in strained graphene, we show here, first via a self-consistent numerical calculations, that the ground state of the Hubbard model in this system is more interesting: while it is locally displaying the expected ferromagnetic ordering,  the sign of the magnetization varies in precisely such a way so that the total space-integrated magnetization in fact vanishes. Since the magnetization is tied to  the zero-modes, however, its support, as well as its sign, also switches between the two sublattices when traversing the system from its bulk to the boundary, so that the N\'{e}el order parameter is also finite, and of the same sign everywhere. The appearance of this  unusual magnetic ordering is also supported by a quantum Monte Carlo calculation on  the Hubbard model on a half-filled  strained honeycomb lattice.  We name this unconventional  magnetic ordering the \emph{global (edge-compensated) antiferromagnet}. Recent experimental progress in realizing the axial magnetic field in real and artificial graphene,\cite{Levy10, Lu12, Gomes12} offers hope for the detection of this unusual ground state.

\begin{figure*}[htb]
\includegraphics[width=5.75cm,height=4.25cm]{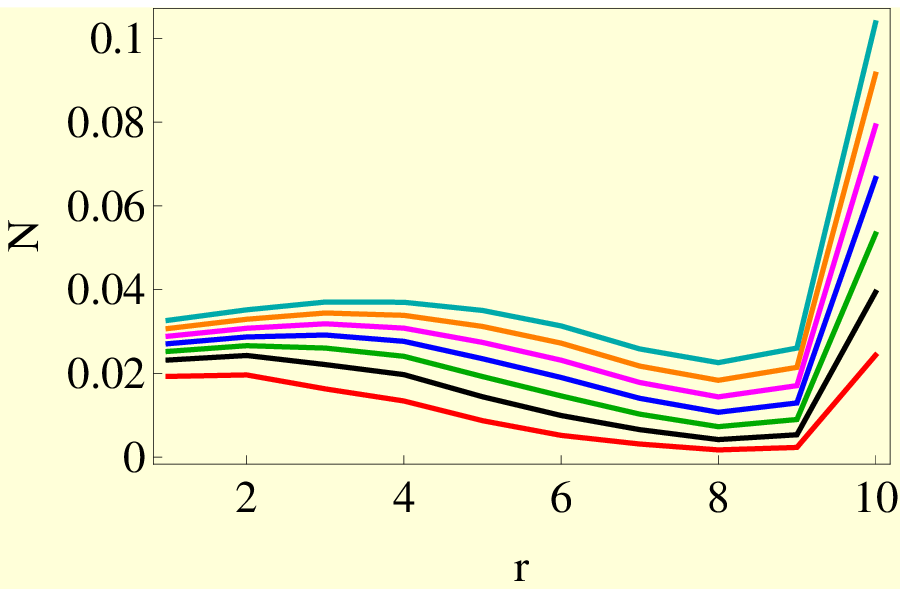}
\includegraphics[width=5.75cm,height=4.25cm]{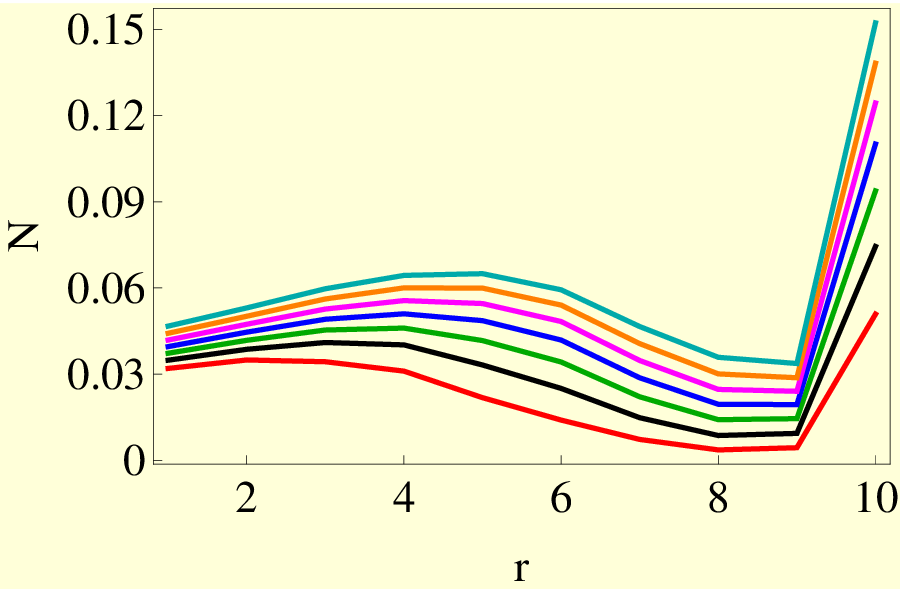}
\includegraphics[width=5.75cm,height=4.25cm]{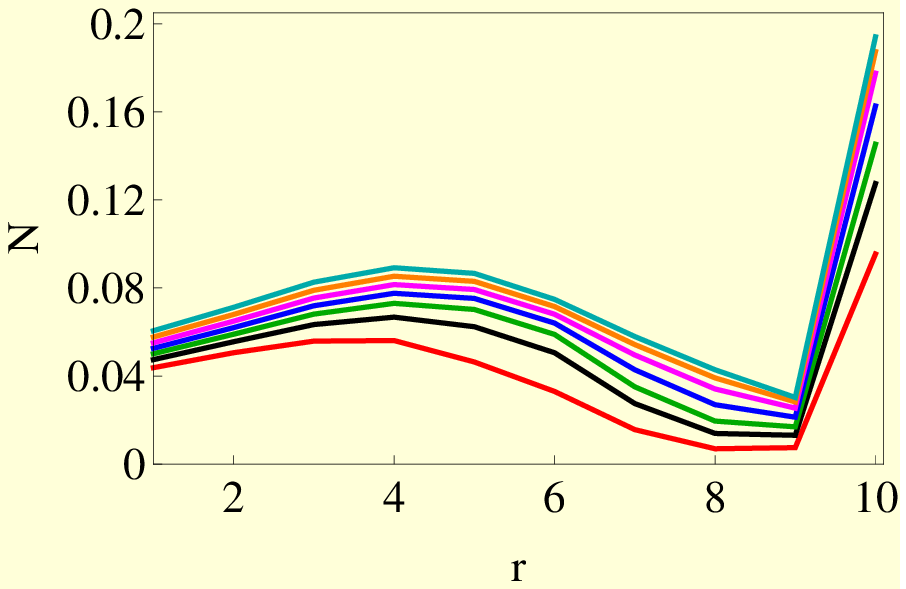}
\includegraphics[width=5.75cm,height=4.25cm]{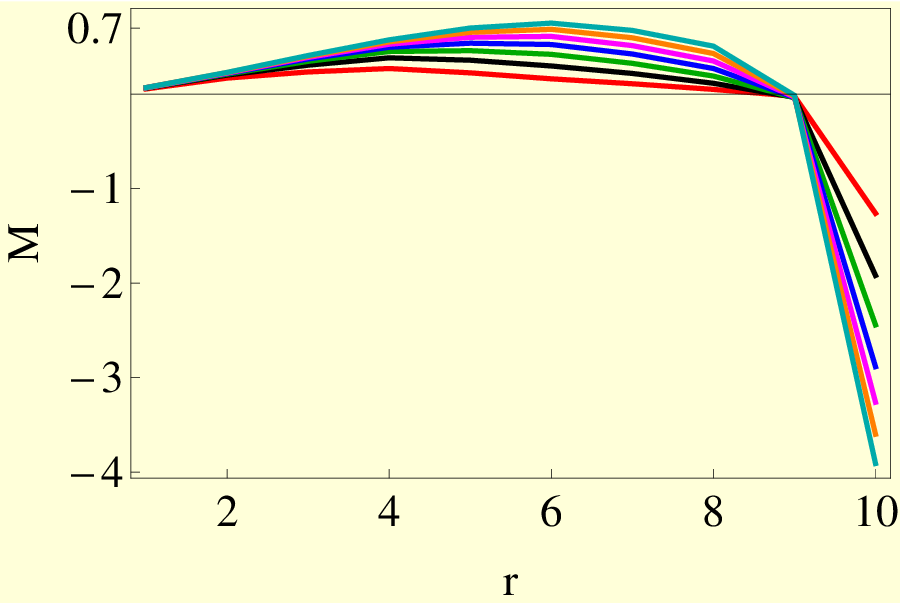}
\includegraphics[width=5.75cm,height=4.25cm]{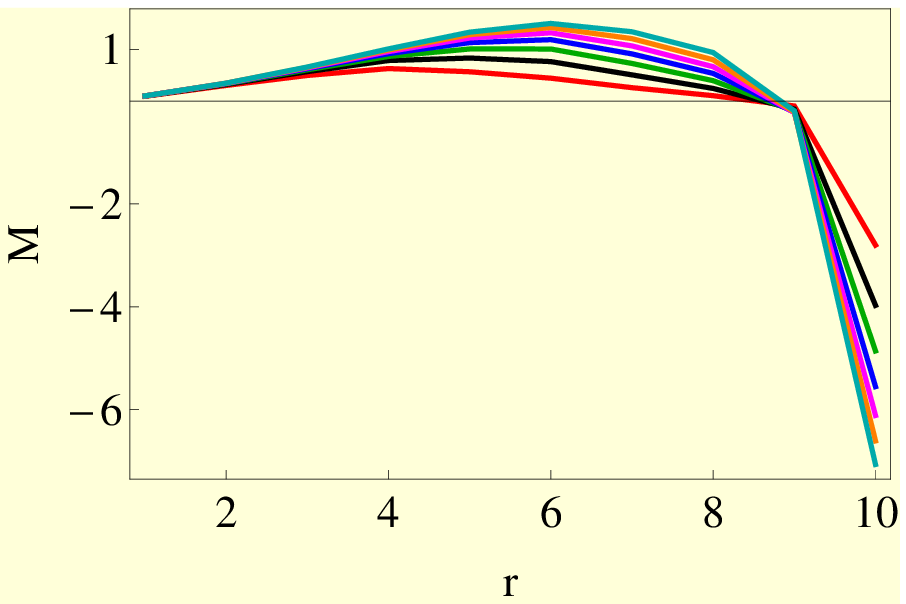}
\includegraphics[width=5.75cm,height=4.25cm]{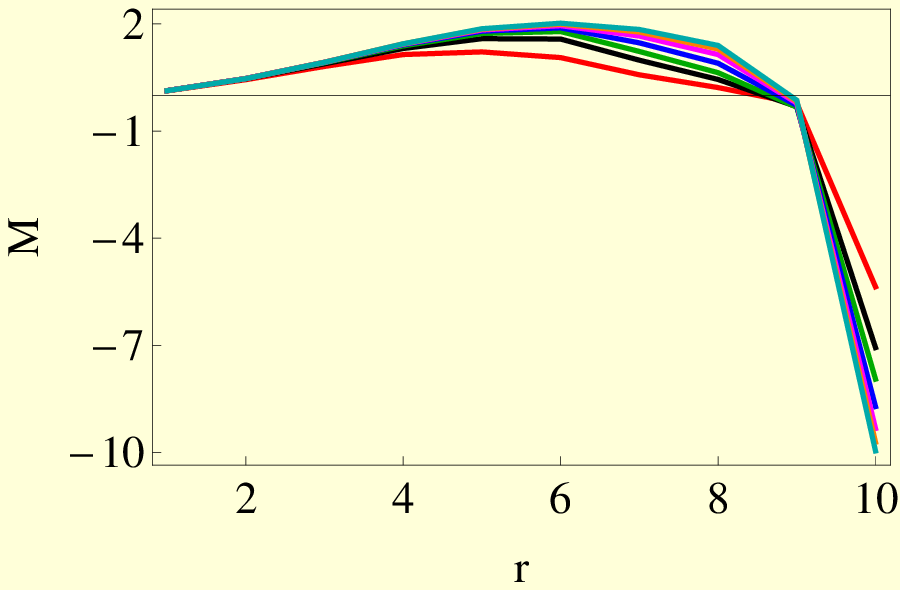}
\caption{(Color online) Upper panel: spatial variation of the local N\'{e}el order parameter ($N$), defined in Eq.~(\ref{AFFMOP}), obtained from the self consistent solution of the magnetic ground state, for various sub-critical onsite Hubbard interaction and roughly uniform axial magnetic fields. The strength of the Hubbard interaction in each plot reads as $U=0.1$ (red), $0.2$ (black),$0.3$ (green), $0.4$(blue), $0.5$ (magenta), $0.6$ (orange), $0.7$(cyan) from bottom to top. Strength of the uniform axial field reads as $b=0.025 \; b_0, 0.035 \; b_0, 0.045 \; b_0$ from left to right, where $b_0=\hbar/(e a^2) \sim 10^4$ T is the axial field associated with the lattice spacing $a \approx 2.5 \mathring{A}$. Lower panel: spatial variation of local ferromagnetic order parameter ($M$) obtained from the same self-consistent solution for the magnetic ground state. Here $N$ is the average N\'{e}el order parameter per unit cell, and $M$ is the total magnetization, in a quasi-circular ring.}
\label{AFFMuniform}
\end{figure*}

The paper is organized as follows. In the next section we propose a specific modulation of the nearest-neighbor hopping amplitude that introduces a finite axial magnetic field in a graphene flake. We also discuss and compare the zero-modes in the presence of the strain-induced axial, and true magnetic fields. In Sec. III we introduce the on-site Hubbard interaction, and discuss possible magnetic ground states in strained graphene. In Sec. IV we present the self-consistent Hatree solution for the magnetic ground state in the presence of (roughly) uniform and well localized axial magnetic fields. We contrast our results with the magnetic ground state in graphene in a true (time-reversal symmetry breaking)  magnetic field in Sec. V. Quantum Monte Carlo simulation of the Hubbard model in strained graphene is presented in Sec. VI. We summarize our findings in Sec. VII.

\section{Axial magnetic field and zero-modes}

Before delving into the effects of the electron-electron interactions, let us set the stage by reviewing the physics of zero-modes in the continuum and on the lattice, and by introducing a specific realization of the axial magnetic fields on a finite honeycomb lattice. As well known, the low energy degrees of freedom in graphene may be collected into an 8-component Dirac spinor $\Psi=\left[ \Psi_\uparrow, \Psi_\downarrow \right]^\top$, where $\Psi^\top_\sigma(\vec{q})=$ $\big[u_\sigma(\vec{K}+\vec{q}),$ $v_\sigma(\vec{K}+\vec{q}),$ $u_\sigma(-\vec{K}+\vec{q}),$ $v_\sigma(-\vec{K}+\vec{q}) \big]$, and $\sigma=\uparrow, \downarrow$ are the two projections of electron spin along the z-axis.\cite{Herbut06,Herbut09} $u_\sigma$ is the fermionic annihilation operator on A-sublattice, generated by the linear combination of basis vectors $\vec{a}_1=(\sqrt{3},-1)a$, $\vec{a}_2=(0,1)a$, where $a$ is the lattice spacing,  and $v^\dagger_\sigma$ is the fermionic creation operator on B-sublattice, where $\vec{B}=\vec{A}+\vec{c}$, with $\vec{c}=(1/\sqrt{3},1)(a/2)$. Two inequivalent Dirac points, or valleys, may be  chosen as at $\pm \vec{K}$, where $\vec{K}=(1,1/\sqrt{3})(2 \pi/a\sqrt{3})$. When $|\vec{q}| \ll |\vec{K}|$, the non-interacting low-energy Hamiltonian with only the nearest-neighbor hopping takes the relativistically invariant form $H_D=\sigma_0 \otimes i \gamma_0 \gamma_j \hat{q}_j$, with the first matrix acting on spin, and the second on sublattice and valley indices.

The coupling of the time-reversal-symmetric axial magnetic field $b(\vec{x})$ to the low-energy Dirac fermions then reads as\cite{Jackiw07, Herbut08, Roy11a,Roy12, Roy13c}
\begin{equation}\label{Diracaxial}
H[\vec{a}]=\sigma_0 \otimes i \gamma_0 \gamma_j \left( q_j - i \gamma_3 \gamma_5 a_j (\vec{x})\right) \equiv M(\chi) H_D M(\chi),
\end{equation}
where $b(\vec{x})=\epsilon_{i j} \partial_i a_j$, and $M(\chi)=\exp{\left[ (\sigma_0 \otimes \gamma_0) \chi(\vec{x})\right]}$. The axial vector potential here is  $a_i (\vec{x})=\epsilon_{i j}\partial_j \chi(\vec{x})$, and therefore $b(\vec{x})=\partial^2 \chi(\vec{x})$. The mutually anticommuting 4-dimensional  $\gamma$-matrices may be represented as $\gamma_0=\sigma_0\otimes\sigma_3$,$\gamma_1=\sigma_3\otimes\sigma_2$,$\gamma_2=\sigma_0\otimes\sigma_1$, $\gamma_3=\sigma_1\otimes\sigma_2$, $\gamma_5=\sigma_2\otimes \sigma_2$. $\left(\sigma_0,\vec{\sigma} \right)$ are the usual two-dimensional Pauli matrices. In this representation the time-reversal symmetry operator is  (anti-linear) $I_K=\sigma_2 \otimes i \gamma_1 \gamma_5 \; K$, where $K$ is the complex conjugation.\cite{Herbut06,Herbut09}

A random distribution of the axial gauge field in graphene, for example, results from the presence of ripples, with the net axial flux as zero. If graphene is deliberately buckled, on the other hand, a finite total flux of the axial field may be introduced, which by index theorem would bring a finite number of states at zero energy. These special zero-modes can be written as
\begin{equation}\label{zerostate}
\Psi_{0,n}\left[\vec{a}\right](\vec{x})\; \propto \; e^{-\chi(\vec{x}) (\sigma_0 \otimes \gamma_0)} \; \Psi_{0,n} \; \left[0\right](\vec{x}).
\end{equation}
The number of zero energy states labeled by index $n$ equals the total axial flux through the system. The matrix $\gamma_0$ in the exponent changes sign between two sublattices, whereas the function $\chi(\vec{x})$ is a monotonic function at a large distance  $|\vec{x}|$ from the location of the axial flux.  The normalizable zero-energy states therefore must reside only on one of the two sublattices, which we will call the sublattice $A$. On the finite honeycomb lattice, these are the bound states in the interior of the system, where the flux is located. The remaining non-normalizable zero energy states, with the support on the sublattice B, on the other hand, in the continuum increase exponentially towards the infinity. On a finite lattice, however, this is tantamount to their  localization {\it near the boundary} of the system.

Eq.~(\ref{Diracaxial}) suggests an introduction of an axial magnetic field on a lattice via the following modification of the nearest-neighbor hopping amplitude ($t$)
\begin{equation}\label{hoppinglattice}
t_{ij} \; \rightarrow \; e^{\chi(i)} \; t \; e^{-\chi(j)},
\end{equation}
where $i \in A$, $j \in B$.\cite{Motrunich02, Roy13} Hereafter we set $t=1$ for convenience. Let us define a quantity, say ${\cal R}$, which counts the minimal number of bonds required to reach a particular site in the system from the central hexagon (centered at $(j_x,j_y)=(0,0)$) in Fig.~1. For all the six sites on the central hexagon of the system ${\cal R}=0$, for example. We then assign the parameter $\chi ({\cal R})$, such that for all the sites with same ${\cal R}$, $\chi ({\cal R})$ is same, as shown in Fig.~1, and the hopping between nearest-neighbor sites are modified according to Eq.~(\ref{hoppinglattice}). Otherwise, $\chi ({\cal R})$ increases monotonically from the center towards the boundary of the system. As a result, a finite axial magnetic field is introduced in the system. For example, if $\chi({\cal R}) \sim {\cal R}^2$ the system experiences a roughly uniform axial field, whereas a bell-shaped localized axial flux around the center of the system can be realized by setting $\chi({\cal R}) \sim \log{\cal R}$. This configuration of $\chi({\cal R})$ is slightly different than in the previous work \cite{Roy13}, with the advantage that the increase of the band width in the presence of axial fields can be somewhat better controlled here. {\color{black} Upon introducing such modification in the nearest-neighbor hopping amplitude, the strained honeycomb lattice, shown in Fig.~1, is invariant under a $C_3$ symmetry, and thus the applied strain in our system is tri-axial.}
\begin{figure*}[htb]
\includegraphics[width=5.75cm,height=4.25cm]{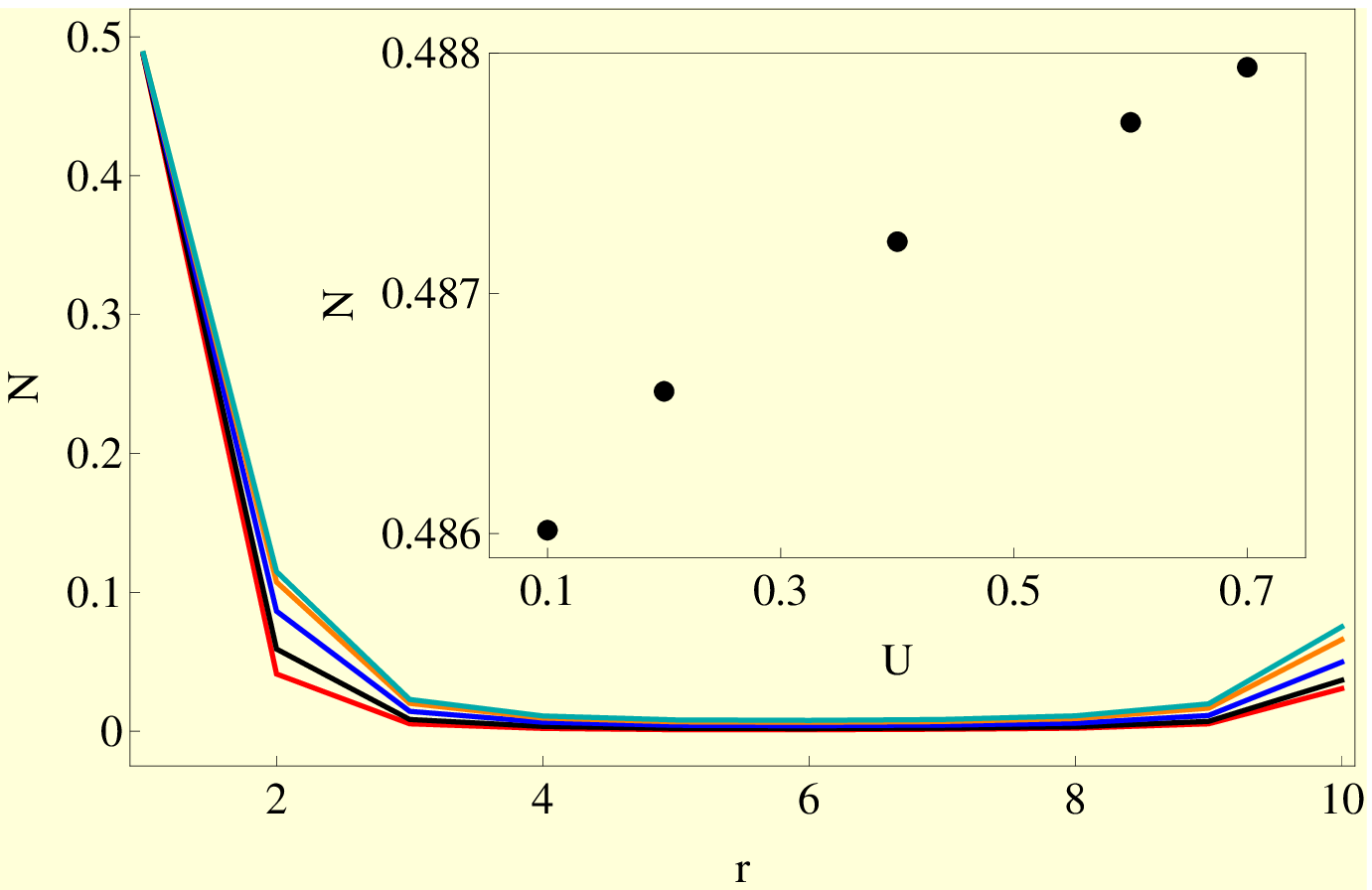}
\includegraphics[width=5.75cm,height=4.25cm]{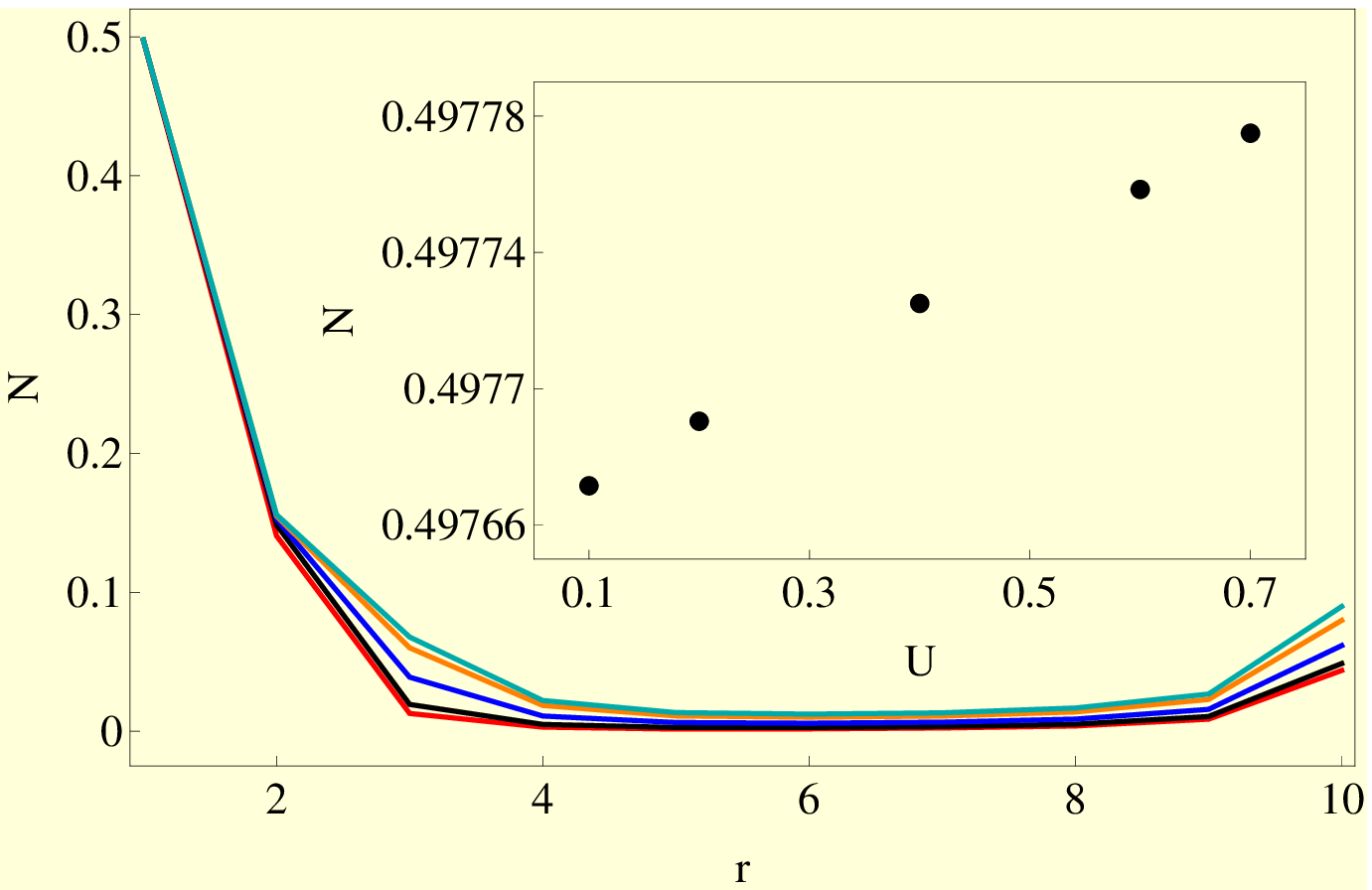}
\includegraphics[width=5.75cm,height=4.25cm]{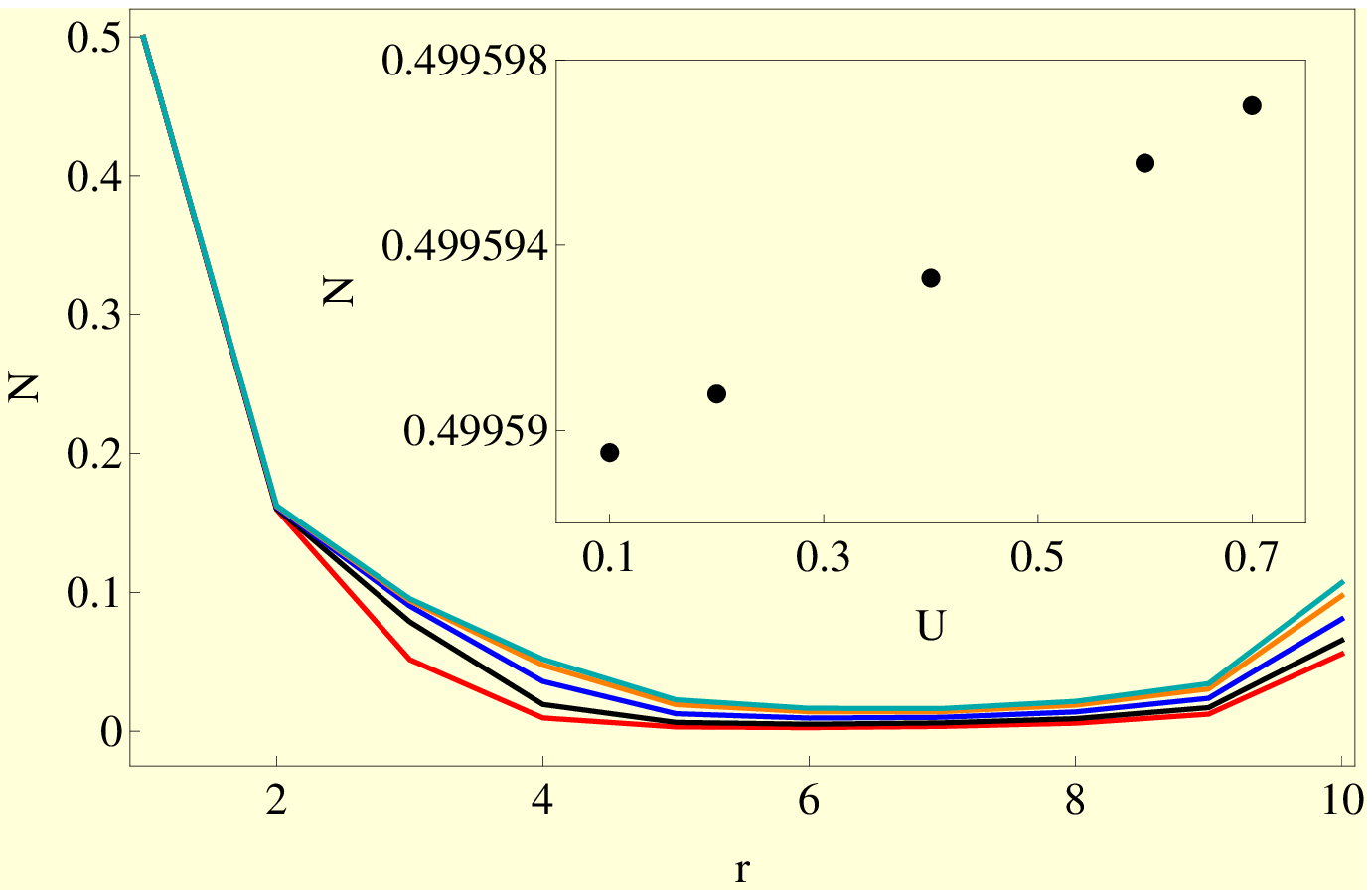}
\includegraphics[width=5.75cm,height=4.25cm]{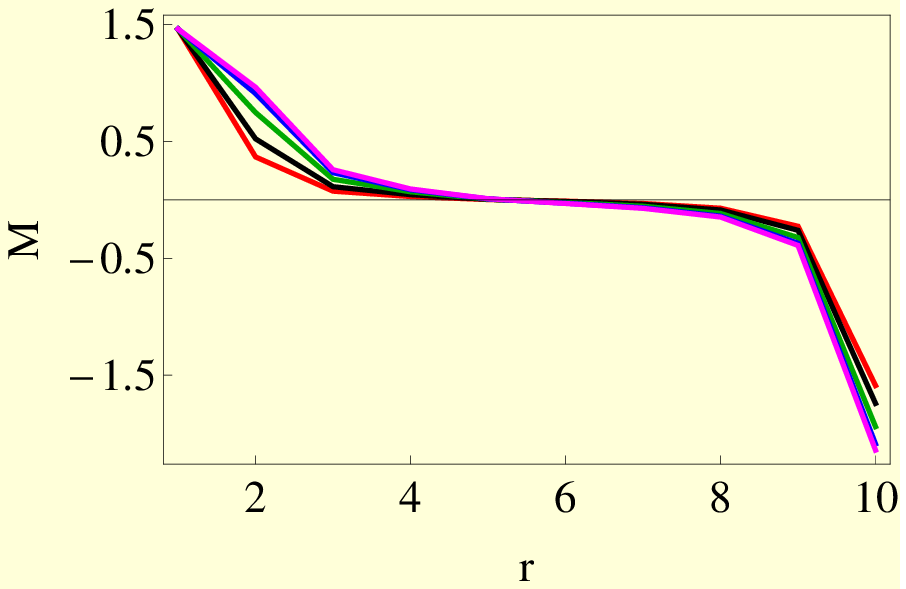}
\includegraphics[width=5.75cm,height=4.25cm]{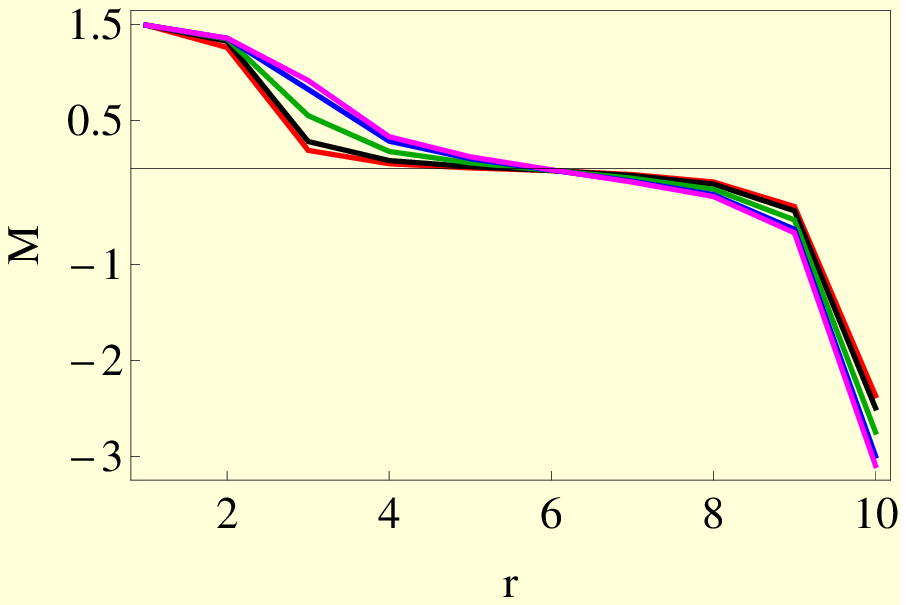}
\includegraphics[width=5.75cm,height=4.25cm]{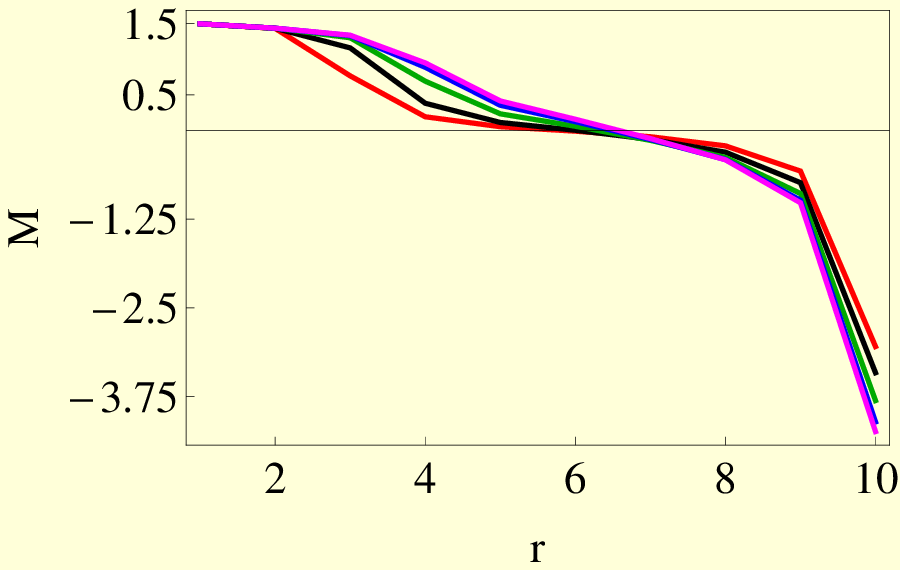}
\caption{(Color online) Upper panel: spatial variation of the local N\'{e}el (N) order parameter, obtained from the self consistent solution of the magnetic ground state, as in Eq.~(\ref{AFFMOP}), for various sub-critical  Hubbard interactions, and bell-shaped localized axial flux peaked around the center of the system. The strength of the Hubbard interaction in each plot reads as $U=0.1$ (red), $0.2$ (black),$0.4$ (blue), $0.6$ (orange), $0.7$ (dark green) from bottom to top. Total flux of the localized axial field reads as $\Phi_{total}=7.85\Phi_0 ,9.42 \Phi_0, 10.99 \Phi_0$ from left to right, where $\Phi_0$ is the flux quanta. Inset(Upper panel): variation of $N(r)$ at the center of the system with the strength of the on-site interaction. Lower panel: spatial variation of local ferromagnet order (M) in the same magnetic ground state for $U=0.1$ (red), $0.2$(black), $0.4$(dark green), $0.6$(blue), $0.7$(magenta). Total flux of the localized axial field in the lower panel is same as in upper one. Here, $N$ and $M$ are same as in Fig.~3.}
\label{AFFMNonuniform}
\end{figure*}

In a finite strained honeycomb lattice there are of course no states at precise zero energy. Nevertheless, irrespective of the spatial profile of the axial magnetic field, there are always a finite number of states in the close vicinity of the zero energy,\cite{Roy13} which bear the signature of the axial magnetic field. Lets consider two such states $| \pm \delta \rangle$, with the energies $\pm \delta$, where $|\delta| \sim 0$. The symmetric $| S \rangle $ and the anti-symmetric $| AS \rangle $ combinations of these two states live on A and B sub-lattice, respectively. In the presence of axial magnetic fields, $| S\rangle$ lives inside the bulk, {\color{black}as shown in Fig.~2(top)} and corresponds to the normalizable zero energy states in Eq.~(\ref{zerostate}). The $|AS \rangle$ lives near the boundary of a finite system, {\color{black}as shown in Fig.2 (bottom)}, and is to be identified as the non-normalizable state in the continuum description. Hence, the bulk and the boundary of a finite graphene system become inequivalent in their sublattice structure in the presence of axial magnetic fields.

{\color{black} Number of recent experiments have already revealed the existence of zero-energy subband in strained real\cite{Levy10, Lu12} and artificial\cite{Gomes12} graphene. In particular, when the axial (uniform) magnetic field $b \sim 60$ T is introduced in strained molecular/artificial graphene, site-resolved scanning tunneling microscope (STM) measurements clearly indicate the existence of zero-energy states on the A-sublattice in the interior of the system\cite{Gomes12}. A careful examination, however, also shows that there are finite number of zero energy states in the close vicinity of the system's boundary, which, on the other hand, are localized on the other sublattice (B). In addition, the zero-modes residing near the boundary are predominantly localized around three of the edges of the hexagonal strained molecular honeycomb lattice, connected through rotations by $2 \pi/3$. All these observations are in accordance with the spatial and sublattice structure of the zero-modes, shown in Fig.~2, obtained by applying a specific modulation of the nearest-neighbor hopping amplitude (Fig.~1) that serves to capture the effect of the axial magnetic field in a finite honeycomb lattice. When $b \leq 45$T, it becomes more difficult to discern the zero-modes on the sublattice B, although near the edge zero bias STM shows bright spots along the nearest-neighbor bonds, possibly revealing this way the overlap among the zero energy states localized on A and B sublattices. Our numerical analysis of the zero modes also exhibits the existence of such region, where the zero modes on two sublattices overlap, see Fig.~2.}

It is worth contrasting the structure of the zero energy subspaces in the presence of true vs. axial magnetic field. Coupling of Dirac fermions to the true magnetic field involves the same  Eq.~(\ref{Diracaxial}), except for the replacement of the matrix  $i\gamma_3 \gamma_5$ by the 4-dimensional unit matrix, and for taking the matrix  $M(\chi)$ to be $ \exp{\left[ (\sigma_0 \otimes i \gamma_3 \gamma_5) \chi(\vec{x})\right]}$. Likewise, the zero energy states in the true magnetic field still have the  form of $\Psi_{0,n}$ in Eq.~(\ref{zerostate}), but with the matrix $\gamma_0$ in the exponent replaced by $i \gamma_{3} \gamma_{5}$. As a consequence, the normalizable zero-modes now reside on both sublattices. In a finite graphene system in a true magnetic field, the near-zero-energy states, residing in the bulk, similarly populate equally both sublattices. The non-normalizable zero-modes in the continuum, located near the boundary in a finite system, are {\it also} shared equally between the two sublattices. Hence, in stark contrast to the axial field, in the presence of true magnetic field, the interior and the boundary of a finite honeycomb lattice have the same unresolved sublattice structure.

\section{Magnetic ground states with Hubbard interaction}

 Let us focus next on the effect of electron-electron interaction in strained graphene, with the chemical potential ($\mu$) tuned to the charge neutrality point. Here we consider only the onsite Hubbard interaction ($U$) among the fermions. The standard Hubbard Hamiltonian is
 \begin{equation}
 H = -t \sum_{\langle i,j \rangle, \sigma} ( u^\dagger _{i,\sigma} v_{j,\sigma} + h. c. ) + H_U,
 \end{equation}
 with the interaction Hamiltonian given by
\begin{equation}
H_U=U \sum_{i}  (  n_{i,\uparrow} -\frac{1}{2} )  ( n_{i,\downarrow}-\frac{1}{2}) - \mu N,
\end{equation}
and where $n_{i,\sigma}$ is the fermionic number operator with spin projection $\sigma=\uparrow, \downarrow$ at site $i$. The total number of electrons in the system is $N$. The charge-neutrality in the system is maintained through the constraint $\mu=0$. The usual Hartree decomposition of the onsite interaction leads to an effective single particle Hamiltonian
\begin{eqnarray}
H_U &=& U \sum_{x=A,B} ( \langle n_{x, \uparrow} \rangle -\frac{1}{2} ) ( n_{x,\downarrow} -\frac{1}{2} )
+( \langle n_{x, \downarrow} \rangle -\frac{1}{2} ) \nonumber \\
&& \times ( n_{x,\uparrow} -\frac{1}{2} ) -\mu N.
\end{eqnarray}
One can rewrite $\langle n_{x, \sigma} \rangle$ as
\begin{equation}\label{decomposition}
\langle n_{A,\sigma}\rangle  =1/2 + \sigma \; \delta_{A,\sigma} (r), \: \langle  n_{B,\sigma} \rangle =1/2 -\sigma \; \delta_{B,\sigma}(r),
\end{equation}
where $\sigma=(+,-)$ respectively represents $(\uparrow,\downarrow)$ projections of electron's spin along $z$-axis. $\delta_{x,\sigma} (r)$ corresponds to the site-dependent local deviation of the electronic density from the uniform background, and is to be determined self-consistently in a finite honeycomb lattice in presence of the axial magnetic field representing strain. We here always take the system to be {\em quasi-circular}, and $r=1,2, \cdots, n$ is a discrete variable, representing the $n^{th}$ ring of around center of the system. In the presence of ordering ($\delta \neq 0$) the overall charge neutrality of the system is achieved through the constraint
\begin{equation}
\sum_{r}\sum_{\sigma=\pm} \bigg[\sigma \delta_{A,\sigma}(r)- \sigma \delta_{B,\sigma}(r) \bigg]=0,
\end{equation}
in addition to $\mu=0$. Notice that the Fock decoupling of the Hubbard term would yield expectation values of the components of the local magnetization that are orthogonal to the chosen ($z$) direction, which we neglect. Depending on the relative sings of the site parameters $\delta$s, one can realize two different magnetic ground states: $(i)$ If the ground state configuration is such that all $\delta_{A/B,\uparrow/\downarrow}>0$ in the above equation, then this would  correspond to an antiferromagnetic phase, whereas $(ii)$ a ground state with $\delta_{A,\uparrow/\downarrow}>0$, but $\delta_{B,\uparrow/\downarrow}<0$ would be identified as a ferromagnetic state. Before we proceed with the numerical simulation of the Hubbard model, it is worth pausing to compare these two magnetic phases in strained graphene.

Magnetization on the two triangular sublattices points in the opposite directions when the ground state is antiferromagnetic, whereas in a ferromagnet the two sublattice magnetizations are aligned. Recall that all the  zero-modes in strained graphene are localized on one sublattice in the bulk, and on the other sublattice near the boundary of a finite system. Therefore, both the N\'{e}el and the ferromagnetic orders in strained graphene give rise to a finite local magnetization everywhere in the system. However, globally the two states may be distinguished.

Due to the spatial separation of the zero-modes that are localized on different sublattices, the magnetization in the state with an equal sign of the N\'{e}el order parameter in the entire system would need to change sign as one approaches the boundary from the bulk, so that the total magnetization could in fact vanish. In the truly ferromagnetic state, on the other hand, the sign of the magnetization in the bulk and the boundary would be the same, so that the system would develop an overall finite magnetization. We will show shortly through a detailed numerical calculation that by taking into account the entire zero subspace in a finite system, the magnetic ground state in strained graphene is uniquely determined to be of the first variety, i. e, a global antiferromagnet with zero total magnetization. Nevertheless, the continuum picture already provides a valuable insight into the nature of the competition between the N\'{e}el  and the ferromagnetic states. The order parameters of these two states read as $\vec{N}=\langle \Psi^\dagger (\vec{\sigma} \otimes \gamma_0) \Psi \rangle$, and $\vec{F}= \langle \Psi^\dagger ( \vec{\sigma} \otimes I_4 ) \Psi \rangle$, respectively. Both states split the zero-energy subspace in strained graphene, and open a gap at the Dirac points. However, the matrix appearing in the N\'{e}el order parameter anticommutes with the Dirac Hamiltonian, and as such it corresponds to a \emph{chiral-symmetry-breaking} mass-term for the Dirac quasi-particles. In the ferromagnetic order parameter, in contrast, the corresponding matrix commutes with the Dirac Hamiltonian. As a result, besides the splitting of the zero-energy subspace common to both orders, the N\'{e}el order at the mean-field level pushes down in energy {\it all} the filled states below the chemical potential. In contrast, the ferromagnetic order parameter splits all the energy levels equally, half up and half down in energy, and therefore only lowers the energy of the half-filled zero-energy subspace. By spontaneously developing the N\'{e}el order the system can thus more efficiently minimize the ground state energy.

\section{Self-consistent calculation}

We now present the results of the self-consistent calculation of the magnetic order parameters with onsite Hubbard interaction ($U$) in strained graphene. We numerically search for the self-consistent solution of $\delta$s with two different initial ansatz for the magnetic ordering $(i)$ when $\delta_{A/B,\uparrow/\downarrow}>0$, which can be idenified as antiferromagnet ordering, and $(ii)$ $\delta_{A,\uparrow/\downarrow}>0, \delta_{B,\uparrow/\downarrow}<0$, which corresponds to a ferromagnetic order. Here, all the $\delta$s are kept as a function of position, and we always maintain the overall charge-neutrality of the system. In the Hartree self-consistent calculation, electronic spin is treated as Ising variable, $\vec{\sigma} \rightarrow \sigma_3$, and the effect of fluctuation is neglected for a moment. Later we treat the fermionic spin as $O(3)$ vectors in a separate quantum Monte Carlo simulation of the Hubbard model in strained graphene, which explicitly takes into account the effect of the fluctuations. We here search for the self-consistent solution of the magnetic order specifically for weak Hubbard interactions, $U \ll U_c$, where $U_c \approx 3.8 t$ is the zero axial field critical strength of the onsite interaction for antiferromagnetic ordering\cite{Assaad13}. Due to the presence of a finite density of state near the zero energy in strained graphene, ordering takes place even for onsite interaction as weak as $U/t=0.1$, irrespective of the spatial profile of the axial magnetic field, and a spectral gap opens up at the Dirac points. The resulting magnetic ground states is insensitive to the initial ansatz $(i)$ or $(ii)$, and thus our self-consistent analysis can be considered as variational calculation. To further explore the nature of such magnetic ground state, we define two local order parameters as
\begin{eqnarray}\label{AFFMOP}
N(r)=\frac{1}{2} \left( \delta_{A,\uparrow} + \delta_{A,\downarrow} + \delta_{B,\uparrow}+\delta_{B,\downarrow}\right)(r), \nonumber \\
M(r)=\frac{1}{2} \left( \delta_{A,\uparrow} + \delta_{A,\downarrow} - \delta_{B,\uparrow}-\delta_{B,\downarrow}\right)(r),
\end{eqnarray}
which correspond to local N\'{e}el and ferromagnetic order parameters, respectively. $\delta$s are as defined in Eq.~(\ref{decomposition}).

We here obtain the numerical results in a quasi-circular honeycomb lattice of $600$ sites or $r_{max}=10$. In such a system the self-consistent solutions for all the $\delta$s are essentially without any finite size effects, for all values of the sub-critical Hubbard interactions, down to $U=0.1$, and for both the uniform and localized axial fields (see the captions of Fig.~3 and Fig.~4 for details of these parameters). Self-consistent solutions of $N(r)$ and $M(r)$ in the presence of roughly uniform axial magnetic fields and a wide range of sub-critical on-site interaction are shown in Fig.~3 (upper panel) and Fig.~3 (lower panel), respectively. From the spatial variation of these two order parameters we see that the local antiferromagnetic  order parameter $N(r)$ is off the same sign in the bulk as well as in the boundary of the system, whereas $M(r)$ near the boundary is of the opposite sign from the bulk. The total magnetization for all chosen values of the interaction and uniform axial fields is \emph{zero} within the numerical accuracy $\sim 10^{-12}$. Therefore, the magnetic ground state is indeed an antiferromagnet, as one would anticipate from the continuum description of this problem. The N\'{e}el order in strained graphene is different from the conventional one on the honeycomb lattice\cite{Assaad13} as it also carries a finite local magnetization everywhere in the system. We dub this unconventional magnetic ground state \emph{global (edge-compensated) antiferromagnet}.

We also obtained the self-consistent solution for the magnetic ground state when the graphene flake is subject to non-uniform axial fields, localized in the vicinity of the center of the system. The spatial variation of the local order parameters $N(r)$ and $M(r)$, for a wide range of sub-critical interactions, and total flux of localized axial magnetic field is shown in Fig.~4 (upper panel), and Fig.~4 (lower panel), respectively, obtained in a quasi-circular system of $600$ sites ($r_{max}=10$). From Figs.~3 and 4, it is clear that the nature of the magnetic ground state is insensitive to the exact profile of the axial field, and it is always the global antiferromagnet. However, the location where the local magnetization changes its sign depends on the profile of the axial field. In the presence of uniform axial field, magnetization flips its sign only very close to the boundary of the system, whereas in the presence of localized fields the same change occurs in the middle of the system. The difference in the position of the domain wall between the two signs of the magnetization is tied to the spatial distribution of the zero energy states in the system. In the  uniform axial field the zero-modes on A-sublattice are distributed over a large portion of the bulk, whereas they are highly localized in the deep interior when the axial field is peaked near the center of the system. Therefore, the local magnetization suffers a change in sign roughly where the zero energy states on A-sublattice loose their support. The magnitude of both of the order parameters $N(r)$ and $M(r)$ increases with the strength of the interaction and of the axial field, as shown in Figs. 2 and 3.

The spatial distribution of the zero-modes also dictates spatial variation of the N\'{e}el order parameter. Note that in the presence of uniform axial fields, the N\'{e}el order parameter $N(r)$ is more or less uniform in the bulk ($r \leq 6$) of the system (see Fig.~3, the upper panel). The abrupt increment in $N(r)$ near the boundary of the system arises from the existence of zero-energy states on the B-sublattice in that region. When the axial field assumes a spatially localized profile, around the center of the system, the N\'{e}el order parameter predominantly develops in the region where the axial flux penetrates the graphene flake; see Fig.~4 (the upper panel). The existence of the zero-energy states on the sites of the B-sublattice near the boundary leads to the spikes in $N(r)$ even when the axial field is inhomogeneous. Therefore, the spatial modulation of the N\'{e}el order follows closely the profile of the axial magnetic field, resembling in this regard the spatial variation of the quantum anomalous Hall insulator in strained graphene with next-nearest-neighbor interaction.\cite{Roy13}

\begin{figure}[htb]
\includegraphics[width=4.25cm,height=3.0cm]{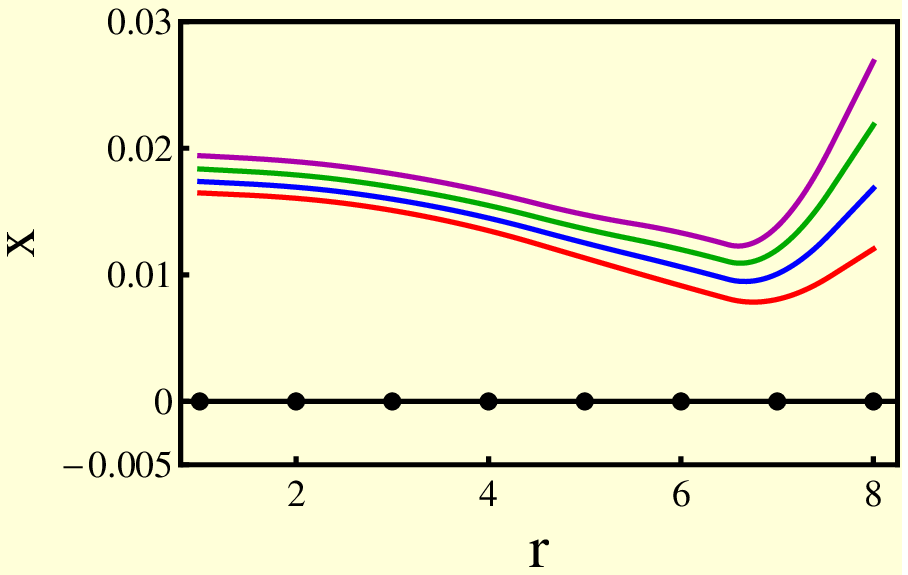}
\includegraphics[width=4.25cm,height=3.0cm]{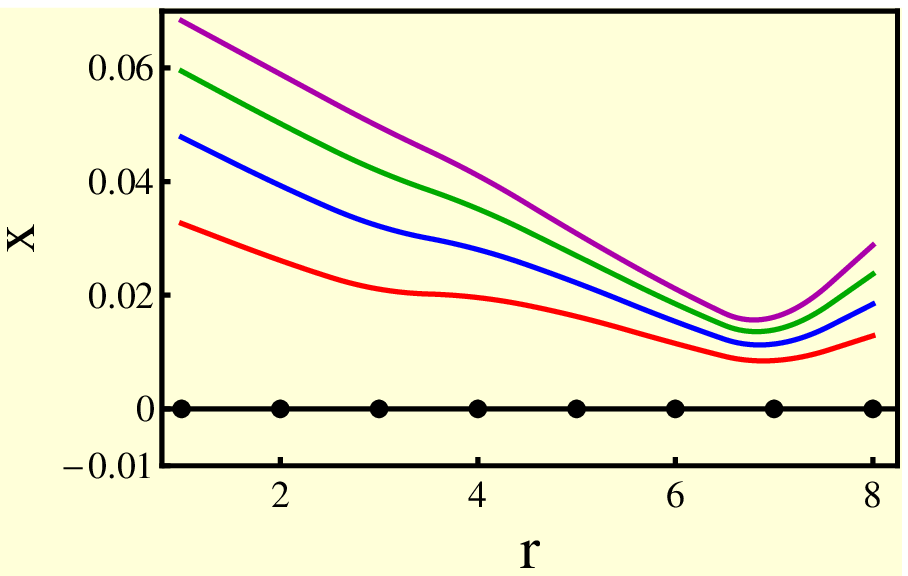}
\caption{(Color online) Left: Spatial variation of the N\'{e}el (in solid lines, with $X=N$) and ferromagnetic (black dots, with $X=M$) order parameters per unit cell, in the presence of uniform true magnetic field of strength $250$ T. Right: The same quantities in a localized true magnetic field, of the strength $650$ T at the center of the system. The average magnetic field at each quasi-circular ring of the honeycomb lattice decreases as $B(r)=(690-40 \; r)$ T, towards the boundary of the system. Strength of the Hubbard interaction reads as $U=0.4$(red), $0.5$(blue), $0.6$(dark green), $0.7$(purple). The ferromagnetic order is identically  \emph{zero} for all values of $U$, both in  uniform and nonuniform true magnetic fields.}
\end{figure}

\section{Hubbard model in true magnetic field}

The conical dispersion of the Dirac quasi-particles quenches into a set of well separated Landau levels, when the graphene flake is subject to a uniform true  magnetic field. Although the discrete quantization of the spectrum smears out if the magnetic field is spatially modulated, a finite number of states at zero energy states always persists, irrespective of the profile of the magnetic fields.\cite{Aharonov79} Finite density of states near the Dirac points, here as well, catalyzes the effect of electron-electron interactions. To compare the present unconventional magnetic ground state of the Hubbard model in strained graphene  with the one in the presence of a true magnetic field, we perform the same numerical self-consistent analysis in a finite honeycomb lattice, placed in uniform and nonuniform true magnetic fields. The orbital effect of the true magnetic field is included by incorporating the Peierls phase into the nearest-neighbor hopping amplitudes.\cite{Roy11b} We here omit the single-particle Zeeman coupling of electron's spin with the magnetic field.\cite{Herbut07,Roy12a} In  true magnetic field the zero energy states are found on both  sublattices in the bulk, as well as near the boundary of the system. In this situation, only the antiferromagnetic ansatz $(i)$ leads to a finite gap at the Dirac point.

\begin{figure*}[htb]
\includegraphics[width=8.75cm,height=6.0cm]{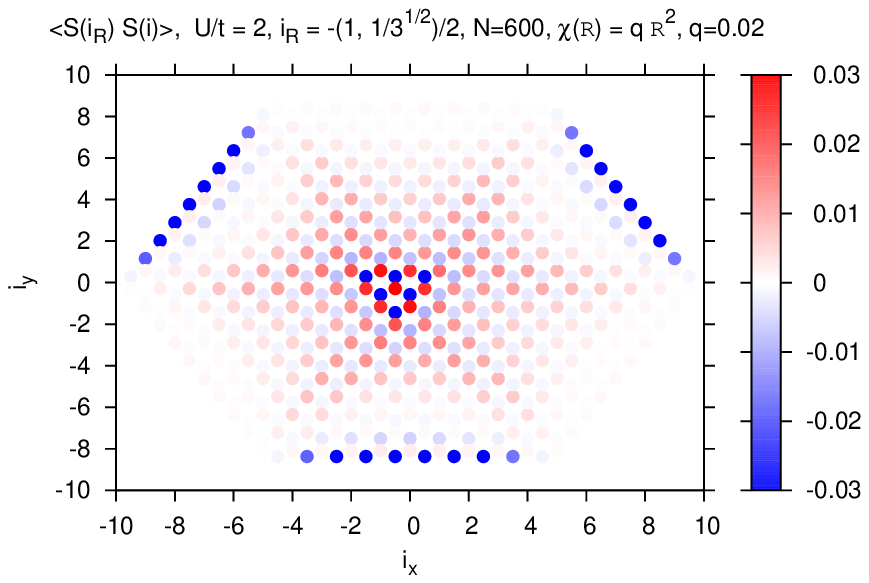}
\includegraphics[width=8.75cm,height=6.0cm]{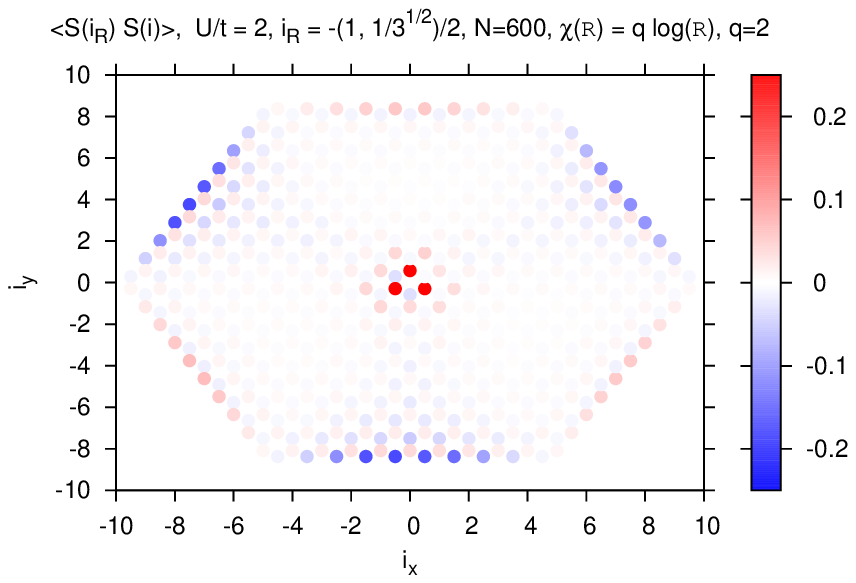} \\
\includegraphics[width=8.75cm,height=6.0cm]{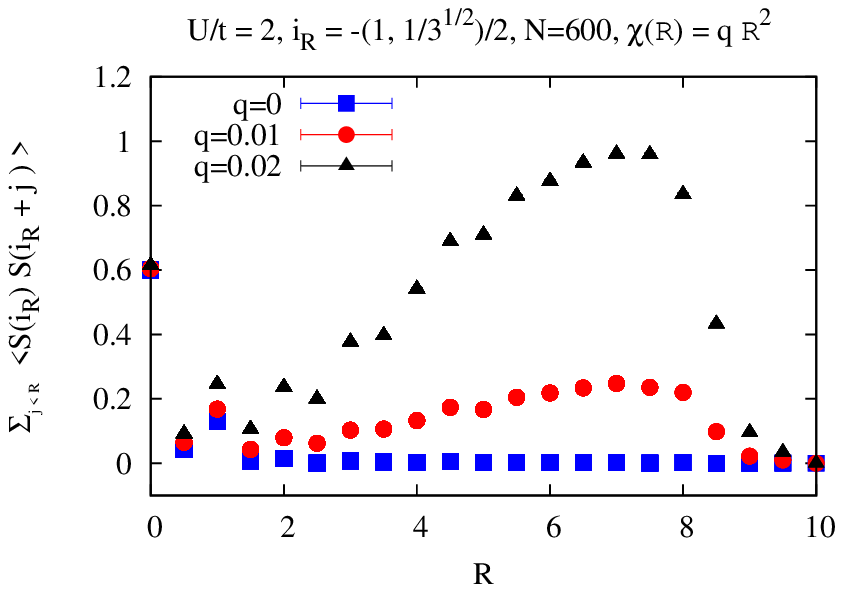}
\includegraphics[width=8.75cm,height=6.0cm]{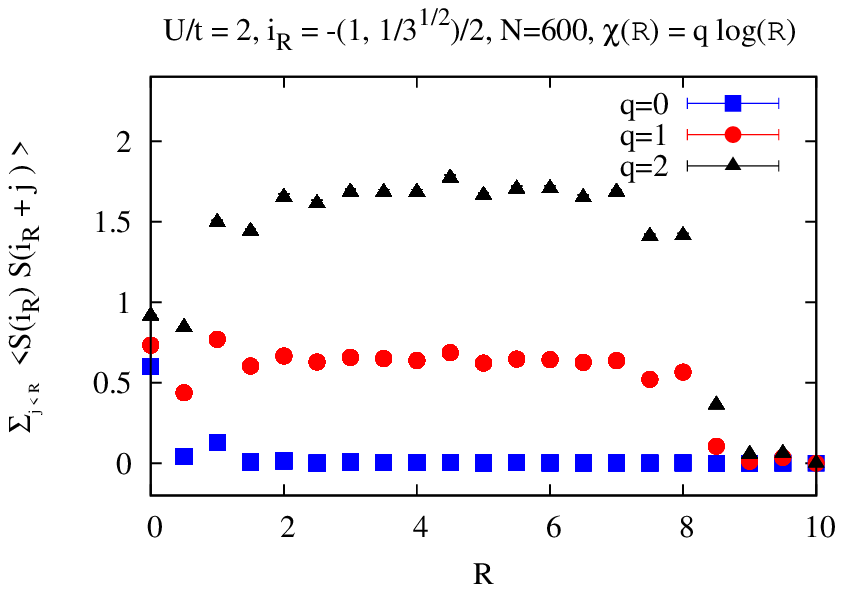}
\caption{(Color online) Quantum Monte Carlo simulations  on an  600 site  flake.   The top panels plot the spin-spin correlation functions  between the  reference point $\pmb{i}_ {\rm R}= (-1/2,-1/2\sqrt{3})$ and  other  sites of the lattice for both the uniform $\chi({\cal R}) = q {\cal R}^2$ and   local $\chi({\cal R}) = q \log( {\cal R})$ axial fields. The two bottom panels plot the integrated spin-spin correlations $S(R)$, defined in Eq.~(\ref{S_R.eq}), exhibiting, at finite values of $q$, the  edge-compensated antiferromagnetic spin structure.}
\label{QMC}
\end{figure*}

From the self consistent solution of the magnetic ground state, obtained in a finite honeycomb system of $384$ sites ($r_{max}=8$), we compute the two local order parameters $N(r)$ and $M(r)$, defined in Eq.~(\ref{AFFMOP}). The results are presented in Fig.~5, when the field is uniform (left) and nonuniform (once again a bell-shaped one, peaked around the center of the system) (right). The local N\'{e}el order parameter $N(r)$ again follows the profile of the true magnetic field, resembling in this regard the spatial distribution of the charge-density-wave order, obtained previously for spinless fermions in honeycomb lattice with the nearest-neighbor interaction.\cite{Roy11b} Due to the existence of a finite density of states at the Dirac point, the antiferromagnetic ordering can be found for the interaction as weak as $U=0.4$.\cite{Herbut07a} On the other hand, $M(r)$ is  \emph{zero} everywhere in the system. The dramatic difference in the magnetic ground states between the  axial and the true magnetic fields arises entirely as a consequence of the distinct structure of the zero-modes. The antiferromagnetic ground state we find in the presence of true magnetic field and at weak interaction ($U \ll U_c$) is the exact replica of the one in graphene at strong interaction ($U \geq 3.8 t$) and in zero field.

\section{Monte Carlo calculation}

\begin{figure*}[htb]
\includegraphics[width=5.7cm]{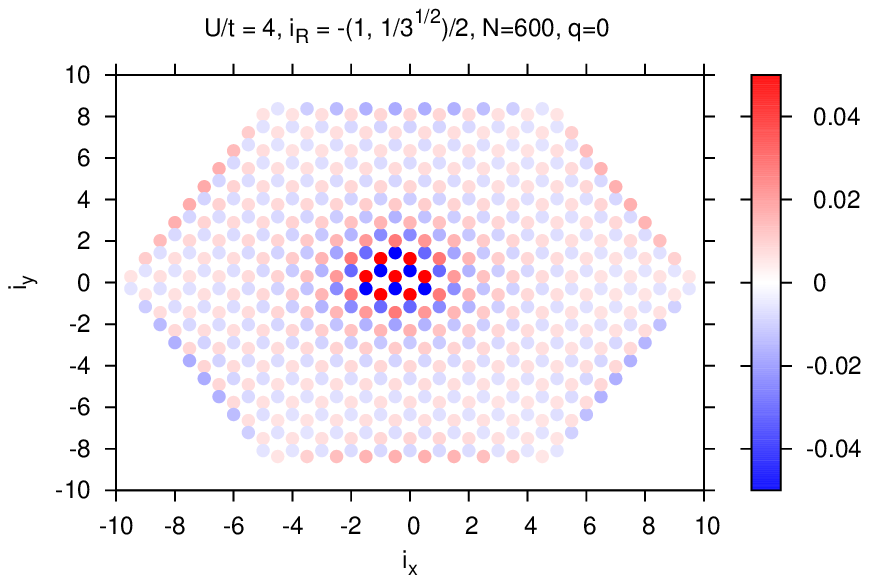}
\includegraphics[width=5.7cm]{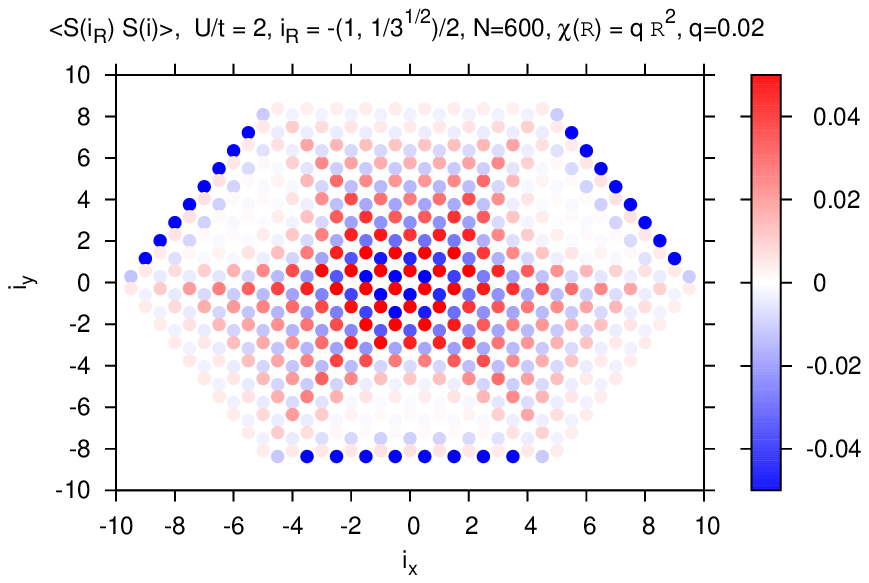}
\includegraphics[width=5.7cm]{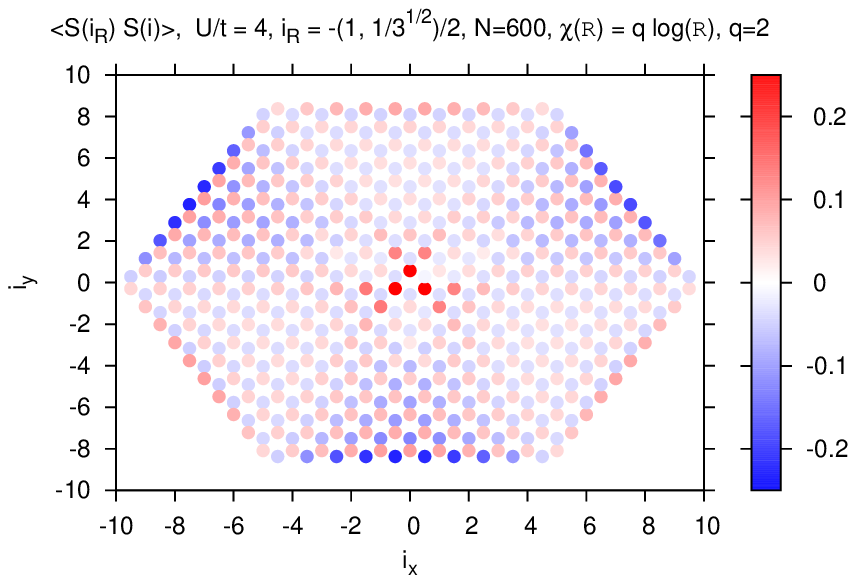} \\
\includegraphics[width=8.75cm,height=6.0cm]{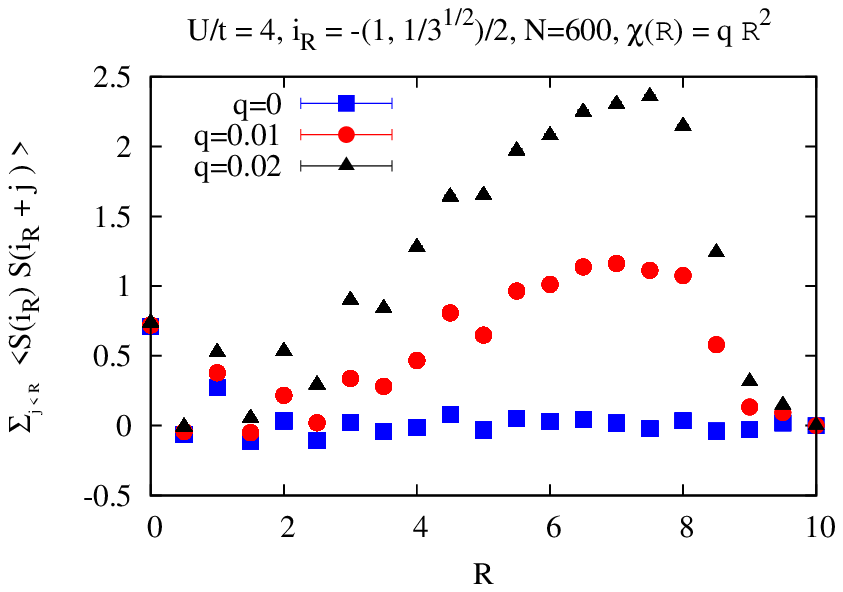}
\includegraphics[width=8.75cm,height=6.0cm]{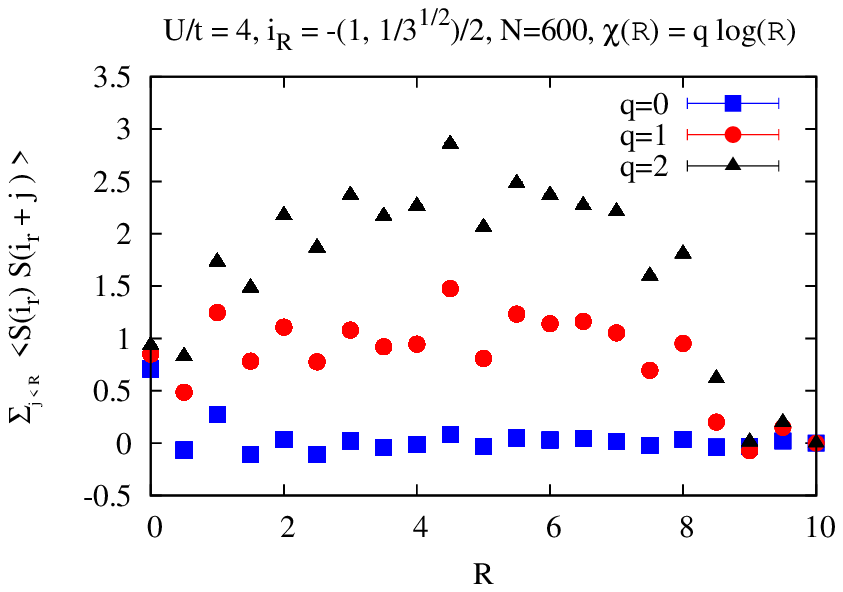}
\caption{ (Color online) Quantum Monte Carlo simulations  on an  600 site  flake. {\color{black}Top panel: the spin-spin correlation functions  between the  reference point $\pmb{i}_ {\rm R}= (-1/2,-1/2\sqrt{3})$ and  other  sites of the lattice in the absence of axial field ($\chi({\cal R})=0$) (left); in the presence of uniform ($\chi({\cal R}) = q {\cal R}^2$) (middle) and local ($\chi({\cal R}) = q \log( {\cal R})$) (right) axial fields.} Integrated spin-spin correlations $S(R)$ (see Eq.~(\ref{S_R.eq})) at $U/t = 4$ for the uniform $\chi({\cal R}) = q {\cal R}^2$ (left) and  local $\chi({\cal R}) = q \log( {\cal R})$ (right) axial fields.}
\label{QMCU4}
\end{figure*}

Axial magnetic fields, resulting  from the  modification of the hopping matrix elements between the nearest neighbors according to Eq.~(\ref{hoppinglattice}), do not  break the particle-hole  symmetry.  Thereby,  auxiliary field quantum  Monte Carlo simulations  do not suffer from the infamous minus sign problem and  accurate simulations on large lattices can be carried out.   Here, we have used the  projective  zero temperature approach based on the equation,
\begin{equation}
  \frac{ \langle \Psi_0  | O  | \Psi_0  \rangle }{ \langle \Psi_0   | \Psi_0  \rangle}   =
\lim_{\Theta \rightarrow \infty}  \frac{ \langle \Psi_T  |  e^{-\Theta H}  O e^{-\Theta H} | \Psi_T \rangle  }{\langle \Psi_T  | e^{-2\Theta H} | \Psi_T \rangle  },
\end{equation}
in which the ground state is filtered out of a single Slater determinant by propagating along the imaginary time axis.  It is beyond the scope of this paper to go into the details of the implementation and the readers are referred to  Ref.~\onlinecite{Assaad08_rev} for an  overview of the algorithm.  Let us  however comment on some  aspects of  our implementation.
The axial field does not break $SU(2)$-spin rotation symmetry and we have found it important to impose  this symmetry by  opting for a discrete Hubbard-Stratonovitch transformation coupling to the charge:
\begin{equation}
	e^{-\Delta \tau  U \left( n_{i} -  1   \right)^2   }  =   \frac{1}{2}\sum_{s=\pm 1} e ^{i \alpha s \left( n_{i} -  1   \right)},
\end{equation}
with $\cos(\alpha) =   e^{-\Delta \tau  U/2}$.    We have  furthermore used a symmetric  Trotter decomposition with  $\Delta \tau t = 0.1$, and  the trial wave function corresponds to the ground state of the non-interacting Hamiltonian.  For this choice of the trial wave function  projection parameters  in the range   $\Theta t = 40-60$ suffice to guarantee  convergence to the ground state within the quoted accuracy.

Since the Monte Carlo simulations do not break  $SU(2)$ spin symmetry  we have to rely on  spin-spin correlations to detect the global edge-compensated antiferromagnetic state. In Fig.~\ref{QMC} (upper panel)  we consider  600-site flakes at  $U/t=2$. This  choice of $U/t$  places us well below  $U_c \approx 3.8 t $ at which the transition to the antiferromagnetic Mott insulating state  occurs in the absence of axial field.\cite{Assaad13}   The reference site $\pmb{i}_ {\rm R}= (-1,-1/\sqrt{3})/2$  is chosen  to belong to the sub-lattice which hosts the
(normalizable)  {\it zero}-energy modes.   In  the very close vicinity of  $\pmb{i}_ {\rm R}$  antiferromagnetic correlations are apparent  and they  rapidly give way to dominant ferromagnetic correlations.
For  the  uniform,   $\chi ({\cal R})  = q {\cal R}^2$, axial field  the extent of the dominant ferromagnetic correlations  is considerably larger than  for the localized one, $\chi({\cal R}) = q \log( {\cal R})$. Finally  strong antiferromagnetic correlations are present at  the edge of the flake.     Hence the overall features present in the quantum Monte Carlo calculations support the mean-field picture of the global edge-compensated antiferromagnetic spin structure.

A  more precise measure for the global edge-compensated antiferromagnetic state can be obtained by  considering
\begin{equation}
	S(R) = \sum_{| {\pmb j}|  < R }   \langle {\pmb S}_{\pmb{i}_{\rm R}} \cdot {\pmb S}_{ \pmb{i}_{\rm R} + \pmb{j}}  \rangle.
\label{S_R.eq}
\end{equation}
Fig.~\ref{QMC} (lower panel) plots this quantity  for the  uniform and  localized  axial fields and at various values of   $q$.    Owing to the singlet nature of the ground state on finite lattices,  $ S(R)   = 0 $, when  $R$ exceeds the radius of the flake.    In the absence of axial field ($q =0$) only short  range antiferromagnetic correlations are present and $S(R)$ quickly decays to zero.   At finite values of the axial field $S(R) $ images the global
edge-compensated antiferromagnetic spin structure.    For the localized field the  zero energy modes are localized around the center
and  $S(R)$  quickly approaches a plateau value before being  compensated by the edge antiferromagnetic correlations. In contrast, for the uniform field $S(R)$  builds up as a function of distance before being again compensated by the edge magnetism.

In Fig. \ref{QMCU4}  we consider larger values of $U/t=4$.  In the absence of the axial field,  this choice of the  Hubbard interaction places us in the antiferromagnetic  Mott insulting phase, {\color{black}shown through the spin-spin correlation in Fig.\ref{QMCU4} (top, left)}. As a consequence, and in comparison to the $U/t=2$ case,  the integrated  spin-spin correlations of Eq.(\ref{S_R.eq}) show  {\it small}  fluctuations up to large  distances.  It is interesting to note that even starting from the Mott insulating state,  the axial field leads to  the same reorganization of the spin-spin correlations as observed at weak couplings, shown in Fig.\ref{QMCU4} (bottom). One will nevertheless observe substantial antiferromagnetic  oscillations superimposed on the edge-compensated antiferromagnetic spin structure. {\color{black} The strain induced restructure of the spin-spin correlation for super-critical interactions ($U/t=4$) is shown in Fig.\ref{QMCU4} (top), for roughly uniform (middle) and localized (right) axial fields.}

\section{Summary and Discussion}

To summarize, we proposed a specific modulation of the nearest-neighbor hopping amplitudes in honeycomb lattice that  captures the coupling of the low-energy Dirac quasi-particles to  the (time-reversal-symmetric) axial magnetic fields. Due to the presence of the  axial magnetic field a finite number of states appears at (near) zero energy, which in turn enhances the effect of electron-electron interaction. Various orderings can take place this way in strained graphene even at weak interactions.\cite{Herbut08, Ghaemi12, Roy13, Roy13a}.

 In this paper we considered only the onsite Hubbard interaction between the fermions, and studied the nature of the magnetic ground state in strained graphene. Due to the special structure of the zero-energy states, which are supported by one sublattice in the bulk of the system, and by the other one near the boundary, the magnetic ground state in strained graphene lacks any analog in pristine graphene, or in graphene in true magnetic fields. Through the numerical self-consistent Hartree  calculation, and a separate quantum Monte Carlo simulation, we established that the magnetic ground state gives rise to both antiferromagnetic and ferromagnetic orders, locally, everywhere in the system. Although the antiferromagnetic order parameter is of the same sign in the entire system, the magnetization changes its sign near the boundary, so that the total magnetization is actually zero. Such ordering takes place even for weak Hubbard interactions. We named the magnetic ground state in strained graphene \emph{edge-compensated antiferromagnet}.

In contrast, the ground state of the Hubbard model on honeycomb lattice subject to true magnetic fields (and with the Zeeman coupling ignored \cite{Herbut07,Roy12a}) is the conventional N\'{e}el antiferromagnet. Through the self-consistent calculation we show that such N\'{e}el ordering takes place again for weak interactions, and the order parameter closely resembles the profile of the true magnetic fields. The magnetization is in this case, however, identically zero everywhere in the system.

{\color{black}
The experimental detection of the global antiferromagnetic phase relies on the measurement of local magnetic moment everywhere in the system, as well as on the existence of the zero-energy states, particularly  near the edge of the system. Recent STM measurements in strained molecular graphene indicate the appearance of zero-modes in the bulk and close to the boundary of the system, which are indeed living on two different sublattices\cite{Gomes12}, in agreement with our numerical analysis (see Fig. 2). 
The local magnetic moment, on the other hand, can be probed by either Magnetic Force Microscope (MFM) or spin-polarized STM measurement.
The latter method successfully established an anti-ferromagnetic ordering on monolayer Fe, resting on tungsten\cite{Kubetzka05}, and revealed the spin structure inside a magnetic vortex core\cite{Wachowiak02}. A systematic measurement of the local magnetic moment using the spin-polarized STM may therefore also detect the unconventional magnetic ground state in strained graphene. The desired suppression of the real-time fluctuations of the local magnetic moment can be obtained either by increasing the system's size, or by increasing the size of the magnetic moments, which, as our analysis would suggest, can be achieved by enhancing the strength of the axial magnetic field and/or the strength of the onsite interaction. The latter can possibly be tuned to a certain degree in the molecular/artificial graphene. Since the local ferromagnetic moment changes its sign roughly where the zero-modes loose their support on the A-sublattice, it is perhaps comparatively easier to detect the proposed global antiferromagnetic phase in strained graphene, when the axial magnetic field is localized near the center of the system. In that situation the ferromagnetic domain wall appears somewhere in the middle of the system (see Fig.~4), whereas, on the other hand, it would lie only very close to the boundary when the axial field is uniform (see Fig.~3).
}

\acknowledgments

B. R. was supported at National High Magnetic Field Laboratory by NSF cooperative agreement No.DMR-0654118, the State of Florida, and the U. S. Department of Energy. FFA is supported by the DFG, under the grant number AS120/9-1 and  AS120/10-1 (Forschergruppe FOR 1807). IFH has been supported by the NSERC of Canada. We thank the  J\"ulich Supercomputing Centre  and the Leibniz-Rechenzentrum in M\"unich for generous allocation of CPU time.


\begin{thebibliography}{10}

\bibitem{Gonzales99}
J.~Gonz\'alez, F.~Guinea, and M.~A.~H. Vozmediano, ``Marginal-Fermi-liquid
  behavior from two-dimensional Coulomb interaction,''
  \href{http://dx.doi.org/10.1103/PhysRevB.59.R2474}{{\em Phys. Rev. B}
  {\bfseries 59} (Jan, 1999) R2474--R2477}.
  \url{http://link.aps.org/doi/10.1103/PhysRevB.59.R2474}.

\bibitem{Sorella92}
S.~Sorella and E.~Tosatti, ``Semi-metal-insulator transition of the Hubbard
  model in the honeycomb lattice,'' {\em Europhys. Lett.} {\bfseries 19} (1992)
  699.

\bibitem{Paiva05}
T.~Paiva, R.~T. Scalettar, W.~Zheng, R.~R.~P. Singh, and J.~Oitmaa,
  ``Ground-state and finite-temperature signatures of quantum phase transitions
  in the half-filled Hubbard model on a honeycomb lattice,'' {\em Phys. Rev. B}
  {\bfseries 72} (2005) 085123.

\bibitem{Khveshchenko01}
D.~V. Khveshchenko, ``Ghost Excitonic Insulator Transition in Layered
  Graphite,'' \href{http://dx.doi.org/10.1103/PhysRevLett.87.246802}{{\em Phys.
  Rev. Lett.} {\bfseries 87} (Nov, 2001) 246802}.
  \url{http://link.aps.org/doi/10.1103/PhysRevLett.87.246802}.

\bibitem{Khveshchenko04}
D.~Khveshchenko and H.~Leal, ``Excitonic instability in layered degenerate
  semimetals,''
  \href{http://dx.doi.org/http://dx.doi.org/10.1016/j.nuclphysb.2004.03.020}{{\em
  Nuclear Physics B} {\bfseries 687} no.~3, (2004) 323 -- 331}.
  \url{http://www.sciencedirect.com/science/article/pii/S0550321304002032}.

\bibitem{Herbut06}
I.~F. Herbut, ``Interactions and phase transitions on graphene's honeycomb
  lattice,'' \href{http://dx.doi.org/10.1103/PhysRevLett.97.146401}{{\em Phys.
  Rev. Lett.} {\bfseries 97} (October, 2006) 146401}.
  \url{http://link.aps.org/doi/10.1103/PhysRevLett.97.146401}.

\bibitem{Herbut09}
I.~F. Herbut, V.~Juri\ifmmode \check{c}\else \v{c}\fi{}i\ifmmode~\acute{c}\else
  \'{c}\fi{}, and B.~Roy, ``Theory of interacting electrons on the honeycomb
  lattice,'' \href{http://dx.doi.org/10.1103/PhysRevB.79.085116}{{\em Phys.
  Rev. B} {\bfseries 79} (Feb, 2009) 085116}.
  \url{http://link.aps.org/doi/10.1103/PhysRevB.79.085116}.

\bibitem{Drut09}
J.~E. Drut and T.~A. L\"ahde, ``Is Graphene in Vacuum an Insulator?,''
  \href{http://dx.doi.org/10.1103/PhysRevLett.102.026802}{{\em Phys. Rev.
  Lett.} {\bfseries 102} (Jan, 2009) 026802}.
  \url{http://link.aps.org/doi/10.1103/PhysRevLett.102.026802}.

\bibitem{Drut09a}
J.~E. Drut and T.~A. L\"ahde, ``Lattice field theory simulations of graphene,''
  \href{http://dx.doi.org/10.1103/PhysRevB.79.165425}{{\em Phys. Rev. B}
  {\bfseries 79} (Apr, 2009) 165425}.
  \url{http://link.aps.org/doi/10.1103/PhysRevB.79.165425}.

\bibitem{Drut09b}
J.~E. Drut and T.~A. L\"ahde, ``Critical exponents of the semimetal-insulator
  transition in graphene: A Monte Carlo study,''
  \href{http://dx.doi.org/10.1103/PhysRevB.79.241405}{{\em Phys. Rev. B}
  {\bfseries 79} (Jun, 2009) 241405}.
  \url{http://link.aps.org/doi/10.1103/PhysRevB.79.241405}.

\bibitem{Castro11}
E.~V. Castro, A.~G. Grushin, B.~Valenzuela, M.~A.~H. Vozmediano, A.~Cortijo,
  and F.~de~Juan, ``Topological Fermi Liquids from Coulomb Interactions in the
  Doped Honeycomb Lattice,''
  \href{http://dx.doi.org/10.1103/PhysRevLett.107.106402}{{\em Phys. Rev.
  Lett.} {\bfseries 107} (Sep, 2011) 106402}.
  \url{http://link.aps.org/doi/10.1103/PhysRevLett.107.106402}.

\bibitem{Grushin13}
A.~G. Grushin, E.~V. Castro, A.~Cortijo, F.~de~Juan, M.~A.~H. Vozmediano, and
  B.~Valenzuela, ``Charge instabilities and topological phases in the extended
  Hubbard model on the honeycomb lattice with enlarged unit cell,''
  \href{http://dx.doi.org/10.1103/PhysRevB.87.085136}{{\em Phys. Rev. B}
  {\bfseries 87} (Feb, 2013) 085136}.
  \url{http://link.aps.org/doi/10.1103/PhysRevB.87.085136}.

\bibitem{Gonzales13}
Gonz\'alez, ``Is graphene on the edge of being a topological insulator?,'' {\em
  Journal of High Energy Physics} {\bfseries 7} (2013) 175.
  \url{http://arxiv.org/abs/1211.3905v1}.

\bibitem{Meng10}
Z.~Y. Meng, T.~C. Lang, S.~Wessel, F.~F. Assaad, and A.~Muramatsu, ``Quantum
  spin liquid emerging in two-dimensional correlated Dirac fermions,''
  \href{http://dx.doi.org/10.1038/nature08942}{{\em Nature} {\bfseries 464}
  no.~7290, (Apr, 2010) 847--851}. \url{http://dx.doi.org/10.1038/nature08942}.

\bibitem{Sorella12}
S.~Sorella, Y.~Otsuka, and S.~Yunoki, ``Absence of a Spin Liquid Phase in the
  Hubbard Model on the Honeycomb Lattice,''
  \href{http://dx.doi.org/http://dx.doi.org/10.1038/srep00992}{{\em Sci. Rep.}
  {\bfseries 2} (2012) 992}.

\bibitem{Assaad13}
F.~F. Assaad and I.~F. Herbut, ``Pinning the Order: The Nature of Quantum
  Criticality in the Hubbard Model on Honeycomb Lattice,''
  \href{http://dx.doi.org/10.1103/PhysRevX.3.031010}{{\em Phys. Rev. X}
  {\bfseries 3} (Aug, 2013) 031010}.
  \url{http://link.aps.org/doi/10.1103/PhysRevX.3.031010}.

\bibitem{Lang13}
T.~C. Lang, Z.~Y. Meng, A.~Muramatsu, S.~Wessel, and F.~F. Assaad, ``Dimerized
  Solids and Resonating Plaquette Order in $\mathrm{SU}(N)$-Dirac Fermions,''
  \href{http://dx.doi.org/10.1103/PhysRevLett.111.066401}{{\em Phys. Rev.
  Lett.} {\bfseries 111} (Aug, 2013) 066401}.
  \url{http://link.aps.org/doi/10.1103/PhysRevLett.111.066401}.

\bibitem{Herbut09a}
I.~F. Herbut, V.~Juri\ifmmode \check{c}\else \v{c}\fi{}i\ifmmode~\acute{c}\else
  \'{c}\fi{}, and O.~Vafek, ``Relativistic Mott criticality in graphene,''
  \href{http://dx.doi.org/10.1103/PhysRevB.80.075432}{{\em Phys. Rev. B}
  {\bfseries 80} (Aug, 2009) 075432}.
  \url{http://link.aps.org/doi/10.1103/PhysRevB.80.075432}.

\bibitem{Roy11}
B.~Roy, ``Multicritical behavior of $ Z_2 \times O(2) $ Gross-Neveu-Yukawa
  theory in graphene,''
  \href{http://dx.doi.org/10.1103/PhysRevB.84.113404}{{\em Phys. Rev. B}
  {\bfseries 84} (Sep, 2011) 113404}.
  \url{http://link.aps.org/doi/10.1103/PhysRevB.84.113404}.

\bibitem{Rosenstein93}
B.~Rosenstein, H.-L. Yu, and A.~Kovner, ``Critical exponents of new
  universality classes,''
  \href{http://dx.doi.org/http://dx.doi.org/10.1016/0370-2693(93)91253-J}{{\em
  Physics Letters B} {\bfseries 314} no.~3–4, (1993) 381 -- 386}.
  \url{http://www.sciencedirect.com/science/article/pii/037026939391253J}.

\bibitem{Kotov12}
V.~N. Kotov, B.~Uchoa, V.~M. Pereira, F.~Guinea, and A.~H. Castro~Neto,
  ``Electron-Electron Interactions in Graphene: Current Status and
  Perspectives,'' \href{http://dx.doi.org/10.1103/RevModPhys.84.1067}{{\em Rev.
  Mod. Phys.} {\bfseries 84} (Jul, 2012) 1067--1125}.
  \url{http://link.aps.org/doi/10.1103/RevModPhys.84.1067}.

\bibitem{Herbut08}
I.~F. Herbut, ``Pseudomagnetic catalysis of the time-reversal symmetry breaking
  in graphene,'' \href{http://dx.doi.org/10.1103/PhysRevB.78.205433}{{\em Phys.
  Rev. B} {\bfseries 78} (Nov, 2008) 205433}.
  \url{http://link.aps.org/doi/10.1103/PhysRevB.78.205433}.

\bibitem{Ghaemi12}
P.~Ghaemi, J.~Cayssol, D.~N. Sheng, and A.~Vishwanath, ``Fractional Topological
  Phases and Broken Time-Reversal Symmetry in Strained Graphene,''
  \href{http://dx.doi.org/10.1103/PhysRevLett.108.266801}{{\em Phys. Rev.
  Lett.} {\bfseries 108} (Jun, 2012) 266801}.
  \url{http://link.aps.org/doi/10.1103/PhysRevLett.108.266801}.

\bibitem{Roy13}
B.~Roy and I.~F. Herbut, ``Topological insulators in strained graphene at weak
  interaction,'' \href{http://dx.doi.org/10.1103/PhysRevB.88.045425}{{\em Phys.
  Rev. B} {\bfseries 88} (Jul, 2013) 045425}.
  \url{http://link.aps.org/doi/10.1103/PhysRevB.88.045425}.

\bibitem{Roy13a}
B.~Roy and V.~Juri\ifmmode \check{c}\else \v{c}\fi{}i\ifmmode~\acute{c}\else
  \'{c}\fi{}, ``Strain-induced time-reversal odd superconductivity in
  graphene,'' {\em arXiv:1309.0507} . \url{http://arxiv.org/abs/1309.0507}.

\bibitem{Guinea10}
F.~Guinea, M.~I. Katsnelson, and A.~K. Geim, ``Energy gaps and a zero-field
  quantum Hall effect in graphene by strain engineering,''
  \href{http://dx.doi.org/10.1038/nphys1420}{{\em Nat Phys} {\bfseries 6}
  no.~1, (Jan, 2010) 30--33}. \url{http://dx.doi.org/10.1038/nphys1420}.

\bibitem{Aharonov79}
Y.~Aharonov and A.~Casher, ``Ground state of a spin-$1/2$ charged particle in a
  two-dimensional magnetic field,''
  \href{http://dx.doi.org/10.1103/PhysRevA.19.2461}{{\em Phys. Rev. A}
  {\bfseries 19} (Jun, 1979) 2461--2462}.
  \url{http://link.aps.org/doi/10.1103/PhysRevA.19.2461}.

\bibitem{Levy10}
N.~Levy, S.~A. Burke, K.~L. Meaker, M.~Panlasigui, A.~Zettl, F.~Guinea,
  A.~H.~C. Neto, and M.~F. Crommie, ``Strain-Induced Pseudo–Magnetic Fields
  Greater Than 300 Tesla in Graphene Nanobubbles,''
  \href{http://dx.doi.org/10.1126/science.1191700}{{\em Science} {\bfseries
  329} no.~5991, (2010) 544--547},
  \href{http://arxiv.org/abs/http://www.sciencemag.org/content/329/5991/544.full.pdf}{{\ttfamily
  http://www.sciencemag.org/content/329/5991/544.full.pdf}}.
  \url{http://www.sciencemag.org/content/329/5991/544.abstract}.

\bibitem{Lu12}
J.~Lu, A.~H.~C. Neto, and K.~P. Loh, ``Transforming moir\'{e} blisters into
  geometric graphene nano-bubbles,''
  \href{http://dx.doi.org/10.1038/ncomms1818}{{\em Nat Commun} {\bfseries 3}
  (May, 2012) 823}. \url{http://dx.doi.org/10.1038/ncomms1818}.

\bibitem{Gomes12}
K.~K. Gomes, W.~Mar, W.~Ko, F.~Guinea, and H.~C. Manoharan, ``Designer Dirac
  fermions and topological phases in molecular graphene,''
  \href{http://dx.doi.org/10.1038/nature10941}{{\em Nature} {\bfseries 483}
  no.~7389, (Mar, 2012) 306--310}. \url{http://dx.doi.org/10.1038/nature10941}.

\bibitem{Jackiw07}
R.~Jackiw and S.-Y. Pi, ``Chiral Gauge Theory for Graphene,''
  \href{http://dx.doi.org/10.1103/PhysRevLett.98.266402}{{\em Phys. Rev. Lett.}
  {\bfseries 98} (Jun, 2007) 266402}.
  \url{http://link.aps.org/doi/10.1103/PhysRevLett.98.266402}.

\bibitem{Roy11a}
B.~Roy, ``Odd integer quantum Hall effect in graphene,''
  \href{http://dx.doi.org/10.1103/PhysRevB.84.035458}{{\em Phys. Rev. B}
  {\bfseries 84} (Jul, 2011) 035458}.
  \url{http://link.aps.org/doi/10.1103/PhysRevB.84.035458}.

\bibitem{Roy12}
B.~Roy, ``Magnetic-field induced inequivalent vortex zero modes in strained
  graphene,'' \href{http://dx.doi.org/10.1103/PhysRevB.85.165453}{{\em Phys.
  Rev. B} {\bfseries 85} (Apr, 2012) 165453}.
  \url{http://link.aps.org/doi/10.1103/PhysRevB.85.165453}.

\bibitem{Roy13c}
B.~Roy, Z.-X. Hu, and K.~Yang, ``Theory of unconventional quantum Hall effect
  in strained graphene,''
  \href{http://dx.doi.org/10.1103/PhysRevB.87.121408}{{\em Phys. Rev. B}
  {\bfseries 87} (Mar, 2013) 121408}.
  \url{http://link.aps.org/doi/10.1103/PhysRevB.87.121408}.

\bibitem{Motrunich02}
O.~Motrunich, K.~Damle, and D.~A. Huse, ``Particle-hole symmetric localization
  in two dimensions,'' \href{http://dx.doi.org/10.1103/PhysRevB.65.064206}{{\em
  Phys. Rev. B} {\bfseries 65} (Jan, 2002) 064206}.
  \url{http://link.aps.org/doi/10.1103/PhysRevB.65.064206}.

\bibitem{Roy11b}
B.~Roy and I.~F. Herbut, ``Inhomogeneous magnetic catalysis on graphene’s
  honeycomb lattice,'' \href{http://dx.doi.org/10.1103/PhysRevB.83.195422}{{\em
  Phys. Rev. B} {\bfseries 83} (May, 2011) 195422}.
  \url{http://link.aps.org/doi/10.1103/PhysRevB.83.195422}.

\bibitem{Herbut07}
I.~F. Herbut, ``SO(3) symmetry between Néel and ferromagnetic order parameters
  for graphene in a magnetic field,''
  \href{http://dx.doi.org/10.1103/PhysRevB.76.085432}{{\em Phys. Rev. B}
  {\bfseries 76} (Aug, 2007) 085432}.
  \url{http://link.aps.org/doi/10.1103/PhysRevB.76.085432}.

\bibitem{Roy12a}
B.~Roy, ``Theory of integer quantum Hall effect in insulating bilayer
  graphene,'' {\em arXiv:1203.6340} . \url{http://arxiv.org/abs/1203.6340}.

\bibitem{Herbut07a}
I.~F. Herbut, ``Theory of integer quantum Hall effect in graphene,''
  \href{http://dx.doi.org/10.1103/PhysRevB.75.165411}{{\em Phys. Rev. B}
  {\bfseries 75} (Apr, 2007) 165411}.
  \url{http://link.aps.org/doi/10.1103/PhysRevB.75.165411}.

\bibitem{Assaad08_rev}
F.~F. {Assaad} and H.~G. {Evertz}, ``{World-line and Determinantal Quantum
  Monte Carlo Methods for Spins, Phonons and Electrons},'' in {\em
  Computational Many Particle Physics}, H.~{Fehske}, R.~{Schneider}, and
  A.~{Wei{\ss}e}, eds., vol.~739 of {\em Lecture Notes in Physics},
  pp.~277--356.
\newblock Springer Verlag, Berlin, 2008.

\bibitem{Kubetzka05}
A.~Kubetzka, P.~Ferriani, M.~Bode, S.~Heinze, G.~Bihlmayer, K.~von Bergmann,
  O.~Pietzsch, S.~Bl\"ugel, and R.~Wiesendanger, ``Revealing Antiferromagnetic
  Order of the Fe Monolayer on W(001): Spin-Polarized Scanning Tunneling
  Microscopy and First-Principles Calculations,''
  \href{http://dx.doi.org/10.1103/PhysRevLett.94.087204}{{\em Phys. Rev. Lett.}
  {\bfseries 94} (Mar, 2005) 087204}.
  \url{http://link.aps.org/doi/10.1103/PhysRevLett.94.087204}.

\bibitem{Wachowiak02}
A.~Wachowiak, J.~Wiebe, M.~Bode, O.~Pietzsch, M.~Morgenstern, and
  R.~Wiesendanger, ``Direct Observation of Internal Spin Structure of Magnetic
  Vortex Cores,'' \href{http://dx.doi.org/10.1126/science.1075302}{{\em
  Science} {\bfseries 298} no.~5593, (2002) 577--580}.
  \url{http://www.sciencemag.org/content/298/5593/577}.

\end{thebibliography}

\providecommand{\href}[2]{#2}\begingroup\raggedright\endgroup
\end{document}